\documentclass[onecolumn,sort&compress,numbers]{els-mrw} 

\usepackage{amsmath,amssymb,amsfonts,amsthm,makeidx,graphicx}
\usepackage{txfonts}
\usepackage{helvet}

\newcommand{\beq}{\begin{equation}}
\newcommand{\eeq}{\end{equation}}

\newcommand{\beqa}{\begin{eqnarray}}
\newcommand{\eeqa}{\end{eqnarray}}

\begin{document}


\chapter{Right-handed neutrinos: seesaw models and signatures}\label{chap1}

\author[1]{Stephen F King}%

\address[1]{\orgname{University of Southampton}, \orgdiv{Department of Physics and Astronomy}, \orgaddress{
Southampton SO17 1BJ, United Kingdom}}


\maketitle

\begin{abstract}[Abstract]
	We give a pedagogical introduction to right-handed neutrinos
as a simple extension to the Standard Model (SM), focussing on seesaw models and their possible experimental signatures. We preface this with a review of the lepton sector of the SM, where charged lepton masses arise from Yukawa couplings and neutrino Majorana masses from the Weinberg operator, leading to a unitary lepton mixing matrix. We first introduce a single right-handed neutrino and the seesaw mechanism, yielding a heavy neutral lepton, then generalise the results to the canonical case of three right-handed neutrinos within a general parameterisation, leading to non-unitary lepton mixing and three heavy neutral leptons, which can detected directly or indirectly via lepton flavour violation or neutrinoless double beta decay. 
We show how the sequential dominance of three right-handed neutrinos with diagonal masses naturally leads 
to an effective two right-handed neutrino model with lepton mixing angle predictions in the constrained cases, but unobservable heavy neutral leptons. On the other hand, with degenerate off-diagonal masses, the two right-handed neutrinos can form a single heavy and observable Dirac neutrino, within a two Higgs doublet or Majoron model. Finally we discuss extra singlet neutrinos which can lead to either a double seesaw or an inverse seesaw, depending on their Majorana masses, where the latter allows observable heavy neutral leptons, and the possibility of a minimal inverse seesaw model where mixing angles can be predicted. 
\end{abstract}

\begin{keywords}
 	Neutrino physics \sep Beyond the Standard Model \sep Right-handed neutrinos \sep Sterile neutrinos \sep Seesaw mechanism
\end{keywords}




\section*{Key Points}

\begin{itemize}
\item Right-handed neutrinos are introduced as a simple and well motivated extension to the Standard Model, leading to the prediction of heavy neutral leptons.
\item Neutrino phenomenology and the Standard Model plus Weinberg operator, leading to neutrino mass and unitary lepton mixing, are briefly reviewed for completeness.
\item The single right-handed neutrino case provides the simplest setting for introducing the key ideas of Majorana mass, the seesaw mechanism, and the phenomenology of a heavy neutral lepton.
\item The three right-handed neutrino paradigm is discussed within a general parameterisation, including non-unitary lepton mixing and the phenomenology of the three heavy neutral leptons, including lepton flavour violation and neutrinoless double beta decay. 
\item The minimal case of two right-handed neutrinos, consistent with neutrino oscillation data, is discussed including either diagonal or off-diagonal right-handed neutrino masses.
\item The case of extra neutrino singlets is considered, focussing on the inverse seesaw mechanism. 
\end{itemize}


\section{Introduction}

The Standard Model (SM) of particle physics~\cite{Glashow:1961tr,Weinberg:1967tq,Salam:1968rm,Glashow:1970gm} involves three families of spin-1/2 chiral fermions, the quarks and leptons, arranged into particular multiplets of the gauge group 
\beq
SU(3)_C\times SU(2)_L\times U(1)_Y
\label{SMgauge}
\eeq
according to 
\beq
\begin{pmatrix}
\nu_{eL} \\
e_L
\end{pmatrix}, \ \ \ \ 
e_R, \ \ \ \
\begin{pmatrix}
u_{L} \\
d_L
\end{pmatrix}^{r,b,g}, \ \ \ \ 
u^{r,b,g}_R, \ \ \ \ 
d^{r,b,g}_R
\label{fam1}
\eeq

\beq
\begin{pmatrix}
\nu_{\mu L} \\
\mu_L
\end{pmatrix}, \ \ \ \ 
\mu_R, \ \ \ \
\begin{pmatrix}
c_{L} \\
s_L
\end{pmatrix}^{r,b,g}, \ \ \ \ 
c^{r,b,g}_R, \ \ \ \ 
s^{r,b,g}_R
\label{fam2}
\eeq

\beq
\begin{pmatrix}
\nu_{\tau L} \\
\tau_L
\end{pmatrix}, \ \ \ \ 
\tau_R, \ \ \ \
\begin{pmatrix}
t_{L} \\
b_L
\end{pmatrix}^{r,b,g}, \ \ \ \ 
t^{r,b,g}_R, \ \ \ \ 
b_R^{r,b,g}
\label{fam3}
\eeq
with respective hypercharges, 
\beq
Y=-\frac{1}{2}, \ \ \ \   -1, \ \ \ \ \ \   \frac{1}{6}, \ \ \ \ \ \ \ \ \ \ \ \ \  \frac{2}{3}, \ \ \ \ \ \ \ \  -\frac{1}{3}
\eeq
plus a complex scalar Higgs doublet $H$ with hypercharge $Y=+1/2$
\beq
H=
\begin{pmatrix}
h^+ \\
h^0
\end{pmatrix}
\label{H}
\eeq
where the vacuum expectation value (VEV) $\langle h^0 \rangle = v/\sqrt{2}$ 
breaks the electroweak symmetry to electromagnetism, resulting in a physical Higgs 
boson~\cite{Higgs:1964pj,Guralnik:1964eu,Englert:1964et,Kibble:1967sv}.
The left-handed (L) chirality fermions form electroweak $SU(2)_L$ doublets, while the 
right-handed (R) chirality fermions form electroweak $SU(2)_L$ singlets. The six quarks 
(u)p, (d)own, (c)harm, (s)trange, (t)op, (b)ottom carry three colours (r)ed, (b)lue, (g)reen under 
$SU(3)_C$. The three families of leptons do not carry colour, and are labelled by the charged lepton mass eigenstates
(e)lectron, muon ($\mu$) and tau ($\tau$). The neutrinos ($\nu$) are leptons with zero electric charge, and
appear only in left-handed doublets together with the charged lepton mass eigenstates, namely electron neutrino ($\nu_{eL}$),
muon neutrino ($\nu_{\mu L}$), tau neutrino ($\nu_{\tau L}$). 
In the SM, these three neutrinos are all massless and are distinguished by separate lepton numbers $L_e$, $L_{\mu}$, $L_{\tau}$. The neutrinos and antineutrinos are distinguished by total lepton number $L=\pm 1$.

The SM survived until almost the end of the last century, but then a series of experiments established the
existence of tiny neutrino masses and rather large neutrino mixing~\cite{Kajita:2016cak,McDonald:2016ixn}. 
In 1998 it was discovered that 
$\nu_{\mu}$ neutrinos, produced from cosmic rays in the upper atmosphere, disappear; later it was shown that this is due to them oscillating between the two identities $\nu_{\mu} - \nu_{\tau}$ on their way to Earth,
which can happen if neutrinos have mass and mixing~\cite{Super-Kamiokande:1998kpq}.
In 2002, the electron neutrinos $\nu_{e}$ from the Sun were shown to also 
disappear, transforming to $\nu_{\mu}$ and $\nu_{\tau}$, which is also a consequence of neutrino oscillations due to their mass and mixing~\cite{SNO:2001kpb}.
Further atmospheric, solar and terrestrial experiments, including reactor experiments, followed, confirming and refining this neutrino oscillation picture~\cite{KamLAND:2002uet,DayaBay:2012fng,RENO:2012mkc,T2K:2011ypd}. 
This culminated in a three family pattern of neutrino masses and mixing angles, and 
we now know that:
\begin{itemize}
\item Neutrinos have tiny masses, approximately one million times smaller than the electron mass.
\item Two of the neutrino masses are similar in mass, unlike the hierarchical charged fermion masses.
\item The neutrino masses break separate lepton numbers $L_e , L_{\mu} , L_{\tau}$ but may or may not respect total lepton number $L=L_e + L_{\mu} + L_{\tau}$,
depending on them being Dirac or Majorana in nature.
\item Neutrinos have large mixing relevant to the atmospheric
and solar experiments, and smaller mixing describing the reactor experiments, comparable to the largest quark mixing angle.
\end{itemize}

\begin{figure}[htb]
\centering
\includegraphics[width=0.4\textwidth]{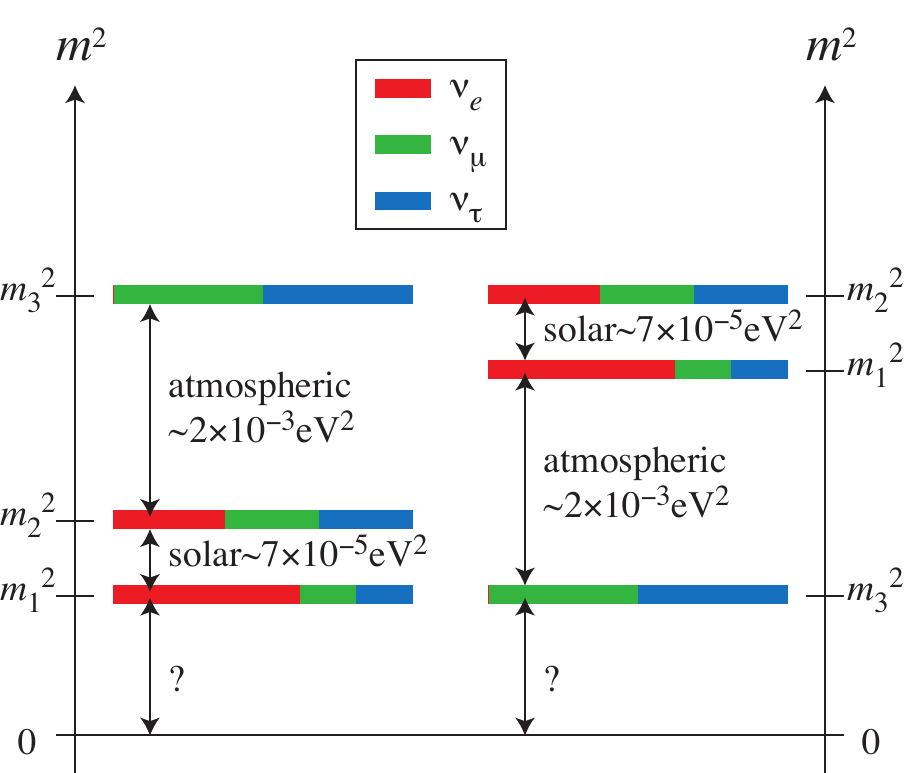}
\caption{\label{mass}{\footnotesize The probability that a particular neutrino
mass state $\nu_i$ with mass $m_i$ contains a particular charged lepton mass basis state 
$(\nu_e, \nu_{\mu}, \nu_{\tau})$ is represented by colours.
The left and right panels of the figure are 
referred to as normal or inverted mass squared ordering, respectively,
referred to as NO or IO.
The value of the lightest neutrino mass is presently unknown.}}
\end{figure}

\begin{figure}[t]
\centering
\includegraphics[width=0.50\textwidth]{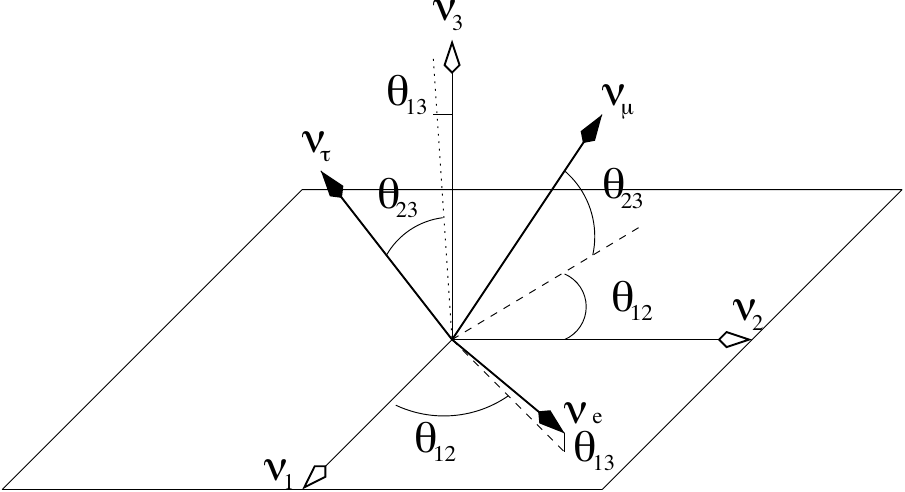}
    \caption{\footnotesize Neutrino mixing angles (assuming zero CP violation) may be represented as Euler angles relating the charged lepton mass basis states $(\nu_e, \nu_{\mu}, \nu_{\tau})$ to the mass eigenstate basis states
    $(\nu_1, \nu_2, \nu_3)$.} \label{angles}
\end{figure}

Neutrino oscillations~\cite{Denton:2025jkt} only depend on the two mass squared differences $\Delta m^2_{21}\equiv m_2^2-m_1^2$, which is constrained by data to be positive, and $\Delta m^2_{31}\equiv m_3^2-m_1^2$, which current data allows to take a positive (normal) or negative (inverted) value, but do not depend on the absolute  neutrino mass scale, as explained in Fig.\ref{mass}.
\footnote{It is common but incorrect 
to refer to the mass squared ordering question as the
``neutrino mass hierarchy''. The ``ordering'' question
is separate from that of whether neutrino masses are
hierarchical in nature or approximately degenerate,
which is to do with the lightest neutrino mass.}
Lepton mixing relevant to the oscillation experiments is 
parameterised by three lepton mixing angles (and one CP violating phase), whose precise definitions will be discussed later, but which may be intuitively understood in Fig.\ref{angles}.
The atmospheric neutrino oscillation data is consistent with bi-maximal $\nu_{\mu}- \nu_{\tau}$ mixing,
with $\theta_{23}\approx 45^{\circ}$~\cite{Super-Kamiokande:1998kpq}, though it could be less than (first octant), or greater than (second octant) this. The solar neutrino oscillation data is consistent with tri-maximal $\nu_e-\nu_{\mu}-\nu_{\tau}$ mixing, with $\theta_{12}\approx 35^{\circ}$~\cite{SNO:2001kpb}. The reactor oscillation experiments determined a third mixing angle 
$\theta_{13}\approx 8.5^{\circ}$~\cite{DayaBay:2012fng,RENO:2012mkc}. The CP violating oscillation phase is currently poorly determined. 
Many neutrino experiments are underway or in the planning stages to address these questions such as T2K~\cite{T2K:2011qtm}, NO${\nu}$A~\cite{NOvA:2004blv}, Daya Bay~\cite{DayaBay:2012fng,DayaBay:2018yms}, JUNO~\cite{JUNO:2015zny}, 
RENO~\cite{RENO:2012mkc,RENO:2018dro}, T2HK~\cite{Hyper-KamiokandeProto-:2015xww}, DUNE~\cite{DUNE:2015lol,DUNE:2020ypp}.
The absolute neutrino mass scale could be determined by measuring the spectrum of electrons emitted from the beta decay of Tritium in the 
Karlsruhe Tritium Neutrino Experiment (KATRIN)~\cite{KATRIN:2019yun,KATRIN:2021uub,Katrin:2024tvg}, where a spectral distortion near the electron end point 
(its maximum energy) would signal neutrino mass. 
The absolute mass of neutrinos is also being probed by many current and planned neutrinoless double beta decay 
experiments~\cite{KamLAND-Zen:2022tow,GERDA:2020xhi,CUORE:2019yfd,EXO-200:2019rkq,nEXO:2021ujk,LEGEND:2021bnm}, which look for a rare kind of double beta decay in which two electrons are always 
emitted at the same time with maximum energy (without any accompanying neutrinos) 
due to the annihilation of the two electron type neutrinos, which would violate lepton number and is only possible if the neutrinos have Majorana mass.

\begin{figure}[t]
\centering
\includegraphics[width=0.4\textwidth]{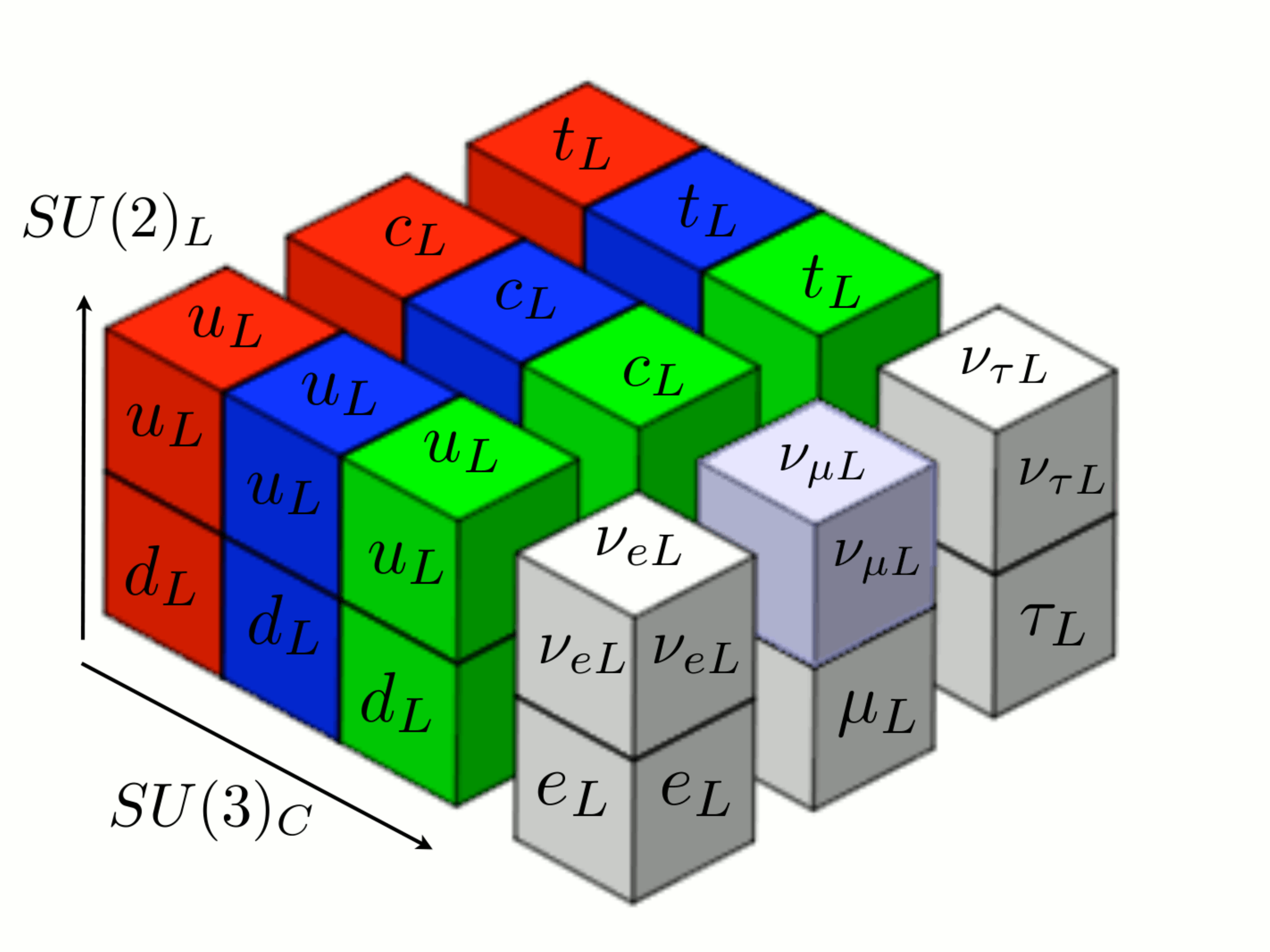}
\includegraphics[width=0.4\textwidth]{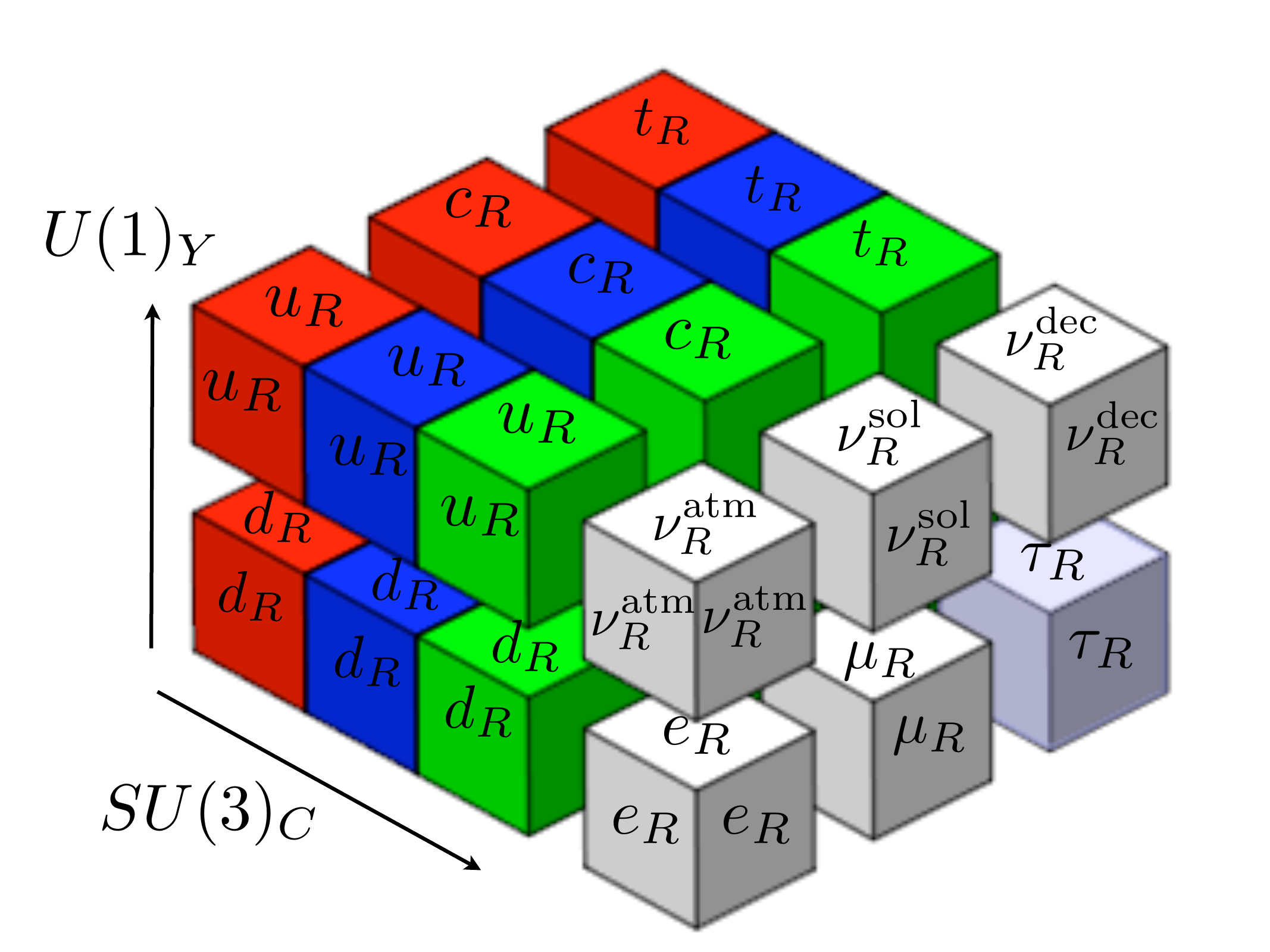}
    \caption{\footnotesize Standard Model $SU(3)_C\times SU(2)_L \times U(1)_Y$ multiplets, 
    left-handed fermions in the left panel, right-handed fermions in the right panel. 
    In addition, we have included three right-handed neutrino singlets. Note that there are then precisely 
    sixteen blocks for each family.} \label{blocks}
\end{figure}

\begin{figure}[t]
\centering
\includegraphics[width=0.50\textwidth]{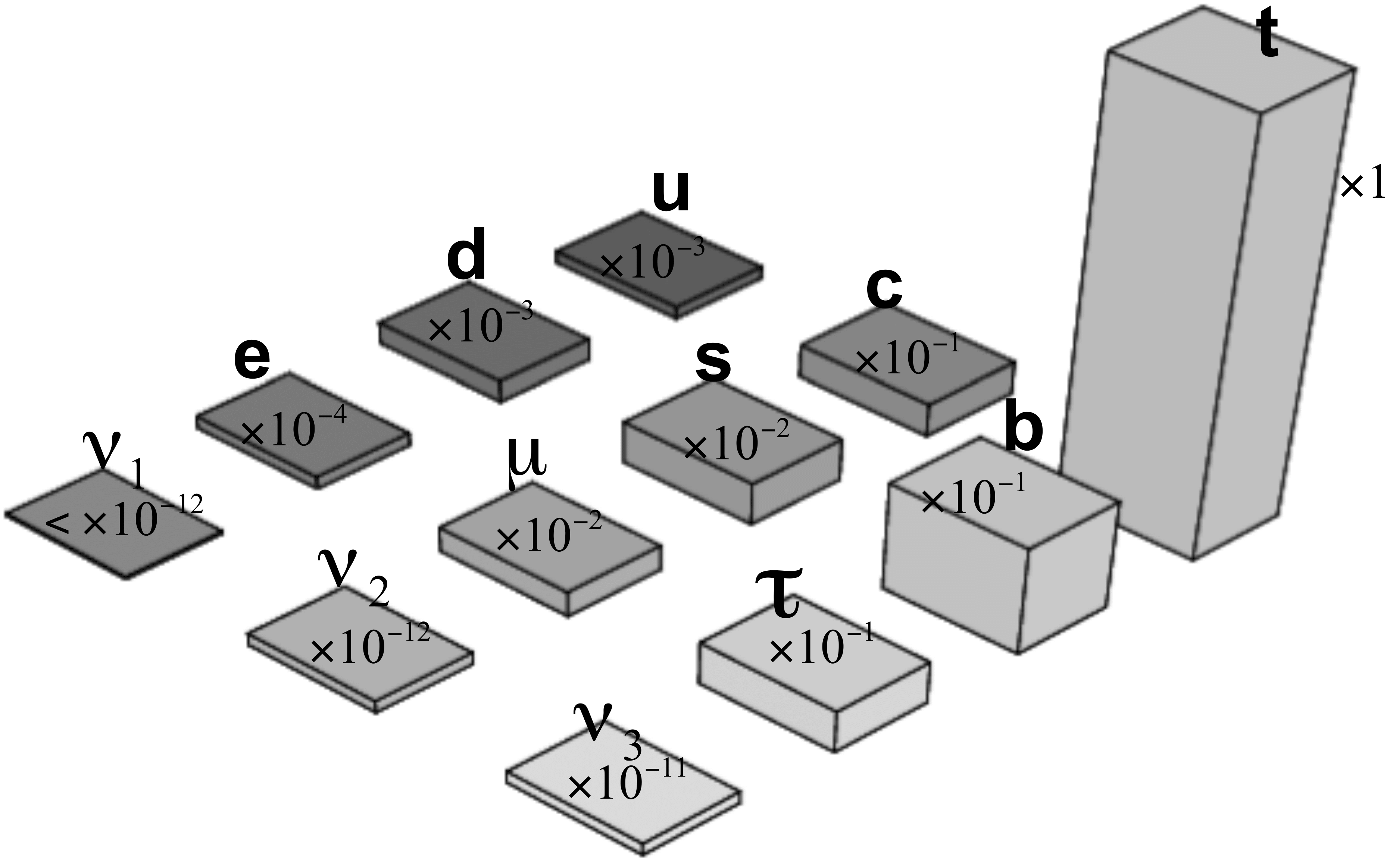}
    \caption{\footnotesize Standard Model mass eigenstate plot where the heights of the towers indicate the masses of the quarks and leptons (true heights need to be scaled by the factors shown). Note the extreme smallness of the neutrino masses. } \label{masses}
\vspace*{-2mm}
\end{figure}

Clearly we must go beyond the SM to understand the origin of neutrino masses.
It is important to note that, in the SM, neutrinos are massless for three reasons:
(1) there are no right-handed neutrinos (RHNs) $\nu_R$; (2) there are only Higgs doublets; 
(3) the SM Lagrangian is renormalisable.
In order to account for neutrino mass and mixing, at least one or more of these conditions must therefore be relaxed.
There are many possibilities for doing this
(see e.g.~\cite{Nath:2006ut,Abada:2007ux,Deppisch:2015qwa,Cai:2017jrq,Cai:2017mow,Agrawal:2021dbo,Han:2022qgg}.)
In this article we shall focus on the possibility and consequences of adding right-handed neutrinos, $\nu_R$, sometimes also called sterile neutrinos since,
as well as being colour and electroweak singlets, they also have zero hypercharge and electric charge.
We are then led to the simple extension of the SM shown in Fig.~\ref{blocks}, where we label the three right-handed neutrinos as $\nu_R^{\rm atm},\nu_R^{\rm sol},\nu_R^{\rm dec}$ which in some models, 
discussed later, are mainly responsible for the physical neutrino masses, $m_3,m_2,m_1$, respectively.
The masses of the quarks and leptons (including neutrinos) are summarised in Fig.~\ref{masses}, which shows
a mysterious pattern not at all understood in the SM. The introduction of RHNs can at least help to explain the smallness of neutrino masses, as we shall discover. 

Interestingly, while some Grand Unified Theories (GUTs) based on $SU(5)$ do not predict RHNs~\cite{Georgi:1974sy},
others do~\cite{Pati:1974yy,Fritzsch:1974nn,Senjanovic:1975rk}.
For example, the SM gauge group in Eq.~\ref{SMgauge} together with baryon minus lepton number, $B-L$,
may be unified into a simple gauge group $SO(10)$~\cite{Fritzsch:1974nn}. The fifteen chiral fermion fields per family in 
Eqs.~\ref{fam1}, \ref{fam2}, \ref{fam3}, when supplemented by one right-handed neutrino field for each family, then make up three
16 dimensional representations of $SO(10)$. Including three right-handed neutrinos, the 16 blocks in Fig.~\ref{blocks} comprising each family may then be glued together into three 16 dimensional families which form complete representations of $SO(10)$
\footnote{It would be nice to be able to glue the three 16 dimensional blocks together to form a single 48 dimensional representation of some family unified gauge group, but such a simple group does not exist.}.
Although the group theory of $SO(10)$ predicts 16 dimensional 
spinorial representations, it does not predict the number of these representations, so the number of families remains a mystery.
However, given that three families of quarks and leptons have been discovered experimentally, then $SO(10)$ GUTs predicts that there should be three right-handed neutrinos to make up the three 16 dimensional representations.
More generally, any $U(1)_{B-L}$ gauge group extension of the SM predicts three RHNs~\cite{King:2004cx,King:2017cwv}.

In Cosmology, right-handed neutrinos can play several possible important roles, which, in order of the most plausible to the most speculative, are as follows.
Firstly, and most plausibly, right-handed neutrinos with Majorana mass around 1 GeV or above could be responsible for the preponderance of matter over antimatter, due to a mechanism known as leptogenesis, which is 
one of the leading candidates for the origin of the baryon asymmetry of the universe (BAU)
\cite{Fukugita:1986hr,Harvey:1990qw,Luty:1992un,Flanz:1994yx,Plumacher:1996kc,Covi:1996wh}~\footnote{the SM by itself yields too small a value for the BAU.}. 
Secondly, right-handed neutrinos could be responsible for dark matter,
since although active neutrino masses of order eV tend to wash out galaxy structures,
since they yield a hot dark matter component, right-handed or sterile 
neutrino masses of order keV can provide acceptable warm dark matter~\cite{Viel:2005qj,Merle:2013gea}.
Thirdly, supersymmetric partners of right-handed neutrinos called sneutrinos
could be responsible for inflating the universe from Planck scale size
to its present size~\cite{Murayama:1992ua,Ellis:2004hy,Antusch:2004hd,Afzal:2024hwj}.
Finally, and most speculatively, right-handed neutrinos may somehow be 
related to dark energy since the eV scale generated via the seesaw mechanism 
happens to be the same order of magnitude as the cosmological constant
(see e.g. \cite{Barbieri:2005gj} and references therein).

The purpose of the present article is to
give a pedagogical introduction to right-handed neutrinos
as a simple extension to the SM, focussing on seesaw models and their possible experimental signatures. Although the scope is limited to right-handed neutrinos, the seesaw methods developed may readily be adapted to other extensions beyond the SM that have been proposed to account for neutrino mass and mixing, as described in detail in the other reviews~\cite{King:2003jb,Strumia:2006db,Altarelli:2010gt,Morisi:2012fg,King:2013eh,King:2014nza,King:2017guk,Feruglio:2019ybq,Xing:2020ijf,Chauhan:2023faf,Ding:2023htn,Ding:2024ozt}. 
The idea is to make the present exposition as accessible as possible to all levels of researchers, both theorists and experimentalists, to act as a stepping stone to the wider literature.

\vspace{0.1in}

The roadmap for the remainder of this article is as follows. 


\begin{itemize}
\item
In section~\ref{leptons} we preface the discussion with an overview of the lepton sector of the SM, and neutrino mass and mixing, starting with the electron and the other charged leptons and neutrinos, showing how the Yukawa couplings to the Higgs doublet generates the charged lepton masses, involving a Yukawa matrix which may be diagonalised without loss of generality to enable the three lepton numbers to be conserved. We then introduce the idea of Majorana masses for the neutrinos and show how the Weinberg operator can provide such masses, leading to neutrino mass and the unitary lepton mixing matrix, whose parameters may be fitted to account for the neutrino oscillation data. This section is independent of any model, and lays the foundation for the remainder of the article.

\item
In section~\ref{SRHN} we introduce a single right-handed neutrino with a large Majorana mass, 
and couplings to the left-handed neutrinos, leading to the seesaw mechanism
for effective light left-handed Majorana masses, capable of describing atmospheric and reactor 
(but not solar) mixing. 
We discuss the resulting heavy neutral lepton, with couplings to the heavy gauge bosons of the SM controlled by a heavy-light mixing angle which, however, is very small in the case of the single right-handed neutrino being responsible for the atmospheric neutrino mass. 

\item
In section~\ref{3RHN} we discuss the canonical case of three right-handed neutrinos, first introducing a general bottom-up parameterisation of the seesaw mechanism, then discussing the non-unitarity of the lepton mixing matrix, and the prediction of three heavy neutral leptons, both controlled by a matrix of heavy-light mixing angles, whose elements may be arbitrarily large in the three right-handed neutrino seesaw case, due to the abundance of free seesaw parameters. We also discuss the indirect effect of heavy neutral leptons on lepton flavour violation and neutrinoless double beta decay.

\item
In section~\ref{2RHN} we focus on the minimal case of two right-handed neutrinos that is consistent with all current neutrino data\footnote{With two right-handed neutrinos the lightest physical neutrino mass is 
predicted to be zero.}, resulting in a considerable reduction of the number of free seesaw parameters.
In sequential dominance, the two right-handed neutrino case emerges as a limiting case of three right-handed neutrinos with diagonal Majorana masses: the dominant right-handed neutrino 
is responsible for the atmospheric neutrino mass and mixing, as in 
the single right-handed neutrino case discussed previously; the subdominant right-handed neutrino provides the solar mixing; a third decoupled right-handed neutrino may be present but plays no significant role in the seesaw mechanism. This leads to a natural neutrino mass hierarchy, with mixing angle predictions possible in constrained cases, but it turns out that the predicted heavy neutral leptons are practically unobservable. 
On the other hand, with degenerate off-diagonal masses, two right-handed neutrinos can form a single heavy and observable Dirac neutrino, due to its asymmetric couplings to lepton doublets, within a two Higgs doublet or Majoron model. 

\item
In section~\ref{extra} we allow extra neutrino singlets (which do not couple to lepton doublets) with either very heavy or very light Majorana masses, together with right-handed neutrinos (which do couple to lepton doublets) which have zero Majorana masses, to begin with, but acquire effective Majorana masses via a double or inverse seesaw mechanism, respectively.
We focus on the inverse seesaw mechanism which allows observable heavy neutral leptons with heavy-light mixing angles depending inversely on the very light singlet Majorana masses. Having increased the number of seesaw parameters, due to the introduction of singlets, we then show how the parameter count may be reduced in a minimal inverse seesaw model, based on sequential dominance, with a constrained model predicting the light neutrino mixing angles while allowing the heavy neutral Dirac leptons to be observable.

\item
Section~\ref{sec:conc} concludes the article.	

\end{itemize}

\section{The Leptons}
\label{leptons}
\subsection{Charged Leptons}
Historically the first known lepton was the electron $e$, discovered by J.J.Thomson in 1897. The relativistic wave equation for the electron was written down by P.A.M. Dirac in 1928 in terms of the four component spinor
\beq
e = 
\begin{pmatrix}
e_L \\
e_R
\end{pmatrix}
\label{e}
\eeq
where $L,R$ represents the left and right chirality states of the spin-1/2 electron field, and we have neglected the electron mass.
For example, $e_L$ is a two component complex Weyl spinor which represents an electron with left-handed chirality, so that its spin vector points in the opposite direction to its momentum vector (classically this would be analogous to a rifle bullet spinning to the left). 
Similarly, $e_R$ is a two component complex Weyl spinor which represents an electron with right-handed chirality.

This simple picture of chiral electrons is strictly only valid for a massless electron, and becomes more complicated once the electron mass term is written down,
\beq
m_e\overline{e}e= m_e(\overline{e}_Le_R + \overline{e}_Re_L)
\label{Dirac}
\eeq
where $m_e\approx 0.511$ MeV.
Such a Dirac mass term connects the $L$ and $R$ degrees of freedom, and allows the electron to flip its chirality.
We then speak of the helicity of the electron being left or right in a particular reference frame. The Dirac mass term in Eq.~\ref{Dirac} respects the gauged $U(1)_{Q}$ symmetry of electrodynamics under which both chiral components of electron field are subject to the same transformation,
\beq
e_{L,R}\rightarrow e^{i \alpha(x)} e_{L,R}
\eeq
where $ \alpha(x)$ is a local spacetime dependent phase. 

In order to explain weak interaction phenomena, S.L. Glashow in 1961~\cite{Glashow:1961tr} introduced an electroweak gauge theory,
now called the Standard Model (SM), based on the non-Abelian gauge group $SU(2)_L$ and the weak hypercharge gauge group $U(1)_Y$
\beq
SU(2)_L \times U(1)_Y
\eeq
under which left-handed chiral electron and neutrino field form a doublet $L_e$ with hypercharge $Y=-1/2$
while the right-handed chiral electron $e_R$ is a singlet with hypercharge $Y=-1$ equal to its electric charge,
\beq
L_e = \begin{pmatrix}
\nu_{eL} \\
e_L
\end{pmatrix}, \ \ \ \ 
e_R
\eeq
It is not possible to write down gauge invariant mass terms, but S. Weinberg in 1967~\cite{Weinberg:1967tq}
and A. Salam in 1968~\cite{Salam:1968rm}
included the complex scalar Higgs\footnote{Actually the complex scalar $SU(2)$ doublet was first discussed by Kibble~\cite{Kibble:1967sv}. Higgs {\it et al}~\cite{Higgs:1964pj,Guralnik:1964eu,Englert:1964et} only considered $U(1)$.} doublet $H$
into the theory with hypercharge $Y=+1/2$, as in Eq.~\ref{H},
\beq
H=
\begin{pmatrix}
h^+ \\
h^0
\end{pmatrix}
\label{H1}
\eeq
where the vacuum expectation value (VEV) $\langle h^0 \rangle = v/\sqrt{2}$ 
breaks the electroweak symmetry to electromagnetism
\beq
SU(2)_L \times U(1)_Y \rightarrow U(1)_{Q}
\eeq
where the electric charge generator $Q$ is given by 
\beq
Q=T_{3L}+Y
\eeq
where $T_{3L}$ is the third generator of $SU(2)_L $ and $Y$ is the hypercharge generator.
The electron mass then arises from the gauge invariant interaction term, 
\beq
y_e \overline{L}_e H e_R + H.c. \rightarrow y_e  \frac{v}{\sqrt{2}} (\overline{e}_Le_R + \overline{e}_Re_L)
\eeq
where we identify the Dirac mass for the electron as, 
\beq
m_e= y_e  \frac{v}{\sqrt{2}}
\eeq
The dimensionless scalar-fermion coupling constant $y_e$ is commonly referred to as a Yukawa coupling after H. Yukawa who studied pion-nucleon couplings in his 1935 paper on strong interactions.

Following the discovery of the muon $\mu$ in 1936 by C.D. Anderson and S. Neddermeyer, and the tau lepton $\tau$ in 1977 by M.L. Perl,
the standard electroweak theory was extended to include three gauge invariant Yukawa terms~\footnote{Of course at the time, Glashow, Weinberg and Salam only knew of the muon terms.}
\beq
y_e \overline{L}_e H e_R + y_{\mu} \overline{L}_{\mu} H \mu_R + y_{\tau} \overline{L}_{\tau} H \tau_R + H.c.
\label{Yuk3}
\eeq
where the three lepton doublets with hypercharge $Y=-1/2$ are 
\beq
L_e = \begin{pmatrix}
\nu_{eL} \\
e_L
\end{pmatrix}, \ \ \ \ 
L_{\mu} = \begin{pmatrix}
\nu_{\mu L} \\
\mu_L
\end{pmatrix}, \ \ \ \ 
L_{\tau} = \begin{pmatrix}
\nu_{\tau L} \\
\tau_L
\end{pmatrix}, \ \ \ \ 
\label{leptondoublets}
\eeq
leading to the three Dirac masses for the charged leptons 
\beq
m_e= y_e  \frac{v}{\sqrt{2}}, \ \ \ \  m_{\mu}= y_{\mu}  \frac{v}{\sqrt{2}}, \ \ \ \ m_{\tau}= y_{\tau}  \frac{v}{\sqrt{2}},
\label{chargedleptonmasses}
\eeq
where $m_e\approx 0.511$ MeV, $m_{\mu}\approx 105.66$ MeV, $m_{\tau}\approx 1777$ MeV
~\cite{ParticleDataGroup:2024cfk}, 
plus three massless left-handed chiral neutrinos $\nu_{eL}, \nu_{\mu L}, \nu_{\tau L}$.
The hierarchy of the three charged lepton masses, 
all of them much smaller than the Higgs VEV $v\approx 246$ GeV,
is parametrised in terms of the three hierarchical Yukawa couplings $y_e \ll y_{\mu} \ll y_{\tau} \ll 1$,
which is unexplained in the SM.

Note that there is an accidental global symmetry which survives electroweak symmetry breaking, namely 
$U(1)_{L_e}$, $U(1)_{L_{\mu}}$, $U(1)_{L_{\tau}}$, associated with the three lepton numbers $L_e$, $L_{\mu}$, $L_{\tau}$, not to be confused with the three lepton doublets of the same name
\footnote{Total lepton number $L= L_e+L_{\mu}+L_{\tau}$ is also conserved.}.
This is simply the symmetry of the three Yukawa terms in Eq.\ref{Yuk3}, where any field with a label $\alpha = e, \mu, \tau$ is associated a global lepton number $L_{\alpha}=1$ under the respective lepton number. For example,
$\nu_{eL}, e_L, e_R$ all have $L_{e}=1$ and $L_{\mu}= L_{\tau}=0$, and so on. The three lepton numbers really express the fact that there are no mass mixing Yukawa terms like for example $\overline{L}_e H \mu_R$ . Of course there is nothing to forbid such mass mixing terms, which are allowed by gauge invariance, but they can be easily rotated away, as we now discuss.

To see this let us write the three lepton doublets as $L_i$ and the three right-handed charged leptons as $e_{Rj}$, where $i,j=1,2,3$, and the Yukawa terms as 
\beq
H \overline{L}_i y^e_{ij}  e_{Rj} + H.c.
\eeq
which may be written as a matrix equation
\beq H
\begin{pmatrix}
 \overline{L}_1 & \overline{L}_2 & \overline{L}_3 
\end{pmatrix}
\begin{pmatrix}
y^e_{11} & y^e_{12} & y^e_{13}\\
y^e_{21} & y^e_{22} & y^e_{23}\\
y^e_{31} & y^e_{32} & y^e_{33}
 \end{pmatrix}
\begin{pmatrix}
e_{R1} \\ e_{R2}  \\ e_{R3}
\end{pmatrix}
+ H.c.
\label{lepmatrix}
\eeq
where the Yukawa matrix can be diagonalised by two independent unitary $3\times 3$ transformations 
$V_{e_L}$ and $V_{e_R}$ acting 
on the triplets of the lepton fields
$L_i$ and $e_{Rj}$,
so the Yukawa matrix can be written in the diagonal basis of Eq.\ref{Yuk3} without loss of generality.
This is achieved by using unitary matrices which satisfy $V^{\dagger} V=I$, where $I$ is the unit matrix, 
with $V^{\dagger}_{e_L} V_{e_L}=I$ and $V^{\dagger}_{e_R} V_{e_R}=I$
inserted before and after the matrix in Eq.\ref{lepmatrix}, with $V_{e_L}$ and $V_{e_R}$ chosen such that,
\beq
V_{e_L} \begin{pmatrix}
y^e_{11} & y^e_{12} & y^e_{13}\\
y^e_{21} & y^e_{22} & y^e_{23}\\
y^e_{31} & y^e_{32} & y^e_{33}
 \end{pmatrix} V^{\dagger}_{e_R}=  
 \begin{pmatrix}
y_e & 0 & 0 \\
0  & y_{\mu} & 0 \\
0 & 0  & y_{\tau}
 \end{pmatrix}
\label{diagchargedleptons1}
 \eeq
where, in the diagonal basis, 
 \beq
  \begin{pmatrix}
 {L}_e \\ {L}_{\mu} \\ {L}_{\tau} 
\end{pmatrix}
= V_{e_L}
  \begin{pmatrix}
 {L}_1 \\ {L}_2 \\ {L}_3 
\end{pmatrix},
\ \ \ \ 
 \begin{pmatrix}
e_{R} \\ \mu_{R}  \\ \tau_{R}
\end{pmatrix}
= V_{e_R}
  \begin{pmatrix}
e_{R1} \\ e_{R2}  \\ e_{R3}
\end{pmatrix}.
\label{diagchargedleptons2}
\eeq

\subsection{Neutrino mass and mixing from the Weinberg operator}

After electroweak symmetry breaking the three chiral neutrinos $\nu_{eL}, \nu_{\mu L}, \nu_{\tau L}$ of the SM 
remain massless. However since the neutrinos are electrically neutral it is possible to write down 
a new kind of mass term envisaged by E. Majorana in 1937~\cite{Majorana:1937vz}, namely 
\begin{equation}
m_{\nu}\overline{\nu_L}\nu_L^{c} 
\label{mL}
\end{equation}
where $\nu_L$ is a left-handed neutrino field and $\nu_L^c$ is its 
CP conjugate field\footnote{Here are elsewhere $\sigma_2$ is the second Pauli matrix.} 
$\nu_L^c=i\sigma_2 \nu_L^*$ (a right-handed antineutrino field). 
Clearly such a Majorana mass term would be forbidden if $\nu_L$ carried electric charge.
However in the SM $\nu_L$ carries hypercharge so it is also forbidden\footnote{Sometimes $\nu_L$ is said to be its own antiparticle, but this is not true since a beam of Majorana neutrinos $\nu_L$ interacts differently from a beam
of antineutrinos $\nu_L^c$ since they have have opposite hypercharge.}. 
The origin of such Majorana masses must lie beyond the SM, but generically they could arise from some high energy gauge invariant non-renormalisable operators as envisaged by S. Weinberg in 1979~\cite{Weinberg:1979sa},
\beq
\kappa_{ij} (L_i^T\tilde{H^*})(\tilde{H}^{\dagger}L_j) \rightarrow m^{\nu \dagger}_{ij}  \overline{\nu^c_{Li}}\nu_{Lj} 
\label{Wop}
\eeq
involving $\tilde{H}=i\sigma_2H^*$, the same Higgs doublet as
in Eq.~\ref{H} but with opposite hypercharge $Y=-1/2$, where $\kappa_{ij}$ are some dimensionful coefficients associated with some high energy scale $\Lambda \gg v$, leading to small Majorana neutrino masses $m^{\nu \dagger}_{ij} = \kappa_{ij}v^2/2\sim v^2/\Lambda$, suppressed by the high energy scale $\Lambda$.

The lepton mass sector below the electroweak symmetry breaking scale is then, in some arbitrary basis
\footnote{Note that the neutrino mass matrix in Eq.~\ref{lepton} is the Hermitian conjugate of that in 
Eq.~\ref{Wop}, where $(m^{\nu})^\dagger=(m^{\nu})^*$, since it is complex symmetric. With this definition the neutrino mass term looks similar to the charged lepton mass term.},
\begin{equation}
	{\cal L}^{\rm lepton}_{\rm mass}  = 
	-\overline e_{Li} m^e_{ij }e_{Rj}  
	-\frac{1}{2}\overline{\nu_{Li}}m^{\nu}_{ij} \nu^c_{Lj} 
	+ {H.c.}
\label{lepton}
\end{equation}
The low energy mass matrices are diagonalised by unitary transformations, as in Eqs.~\ref{diagchargedleptons1}, \ref{diagchargedleptons2},
\begin{eqnarray}
V_{e_L}m^e V_{e_R}^{\dagger}=
\left(\begin{array}{ccc}
m_e&0&0\\
0&m_{\mu}&0\\
0&0&m_{\tau}
\end{array}\right), \ \ \ \ 
V_{{\nu}_L}m^{\nu}V_{{\nu}_L}^{T}=
\left(\begin{array}{ccc}
m_1&0&0\\
0&m_2&0\\
0&0&m_3
\end{array}\right).
\label{diagonalisation}
\end{eqnarray}
The couplings to $W^-$ are given by 
$-\frac{g}{\sqrt{2}}W_{\mu}^-\overline{l}_L\gamma^{\mu}{\nu}_{lL}$, where $l=e, \mu , \tau$ are the charged lepton mass eigenstates, hence
the charged currents in terms of the light neutrino mass eigenstates $\nu_1, \nu_2, \nu_3$ are, 
\begin{eqnarray}
{\cal L}^{CC}_{\rm lepton}= -\frac{g}{\sqrt{2}}W_{\mu}^-
\left(\begin{array}{ccc}
\overline{e}_L  & \overline{\mu}_L &  \overline{\tau}_L
\end{array}\right)
\gamma^{\mu}
U_{\rm PMNS}
\left(\begin{array}{c}
{\nu}_{1} \\ 
{\nu}_{2} \\ 
{\nu}_{3}
\end{array}\right)+H.c.
\label{CC}
\end{eqnarray}
where we have identified the unitary Pontecorvo-Maki-Nakagawa-Sakata (MNS)~\cite{Maki:1962mu} matrix as,
\begin{equation}
\label{enu}
U_{\rm PMNS}=V_{e_L}V_{{\nu}_{L}}^{\dagger}.
\end{equation}
Three of the six phases can be removed since each of the three charged lepton mass terms such as 
$m_e\overline e_{\mathrm{L}}e_{\mathrm{R}}$,
etc., is left unchanged by global phase rotations $e_{\mathrm{L}}\rightarrow e^{i\phi_e}e_{\mathrm{L}}$
and $e_{\mathrm{R}}\rightarrow e^{i\phi_e}e_{\mathrm{R}}$, etc., where the three phases $\phi_e$, etc., are chosen to
leave three physical (irremovable) phases in $U_{\rm PMNS}$.
There is no such phase freedom in the Majorana mass terms 
$-\frac{1}{2} {m_i} \overline{{\nu}_{i}} {\nu}_{i}^{c}$ where $m_i$ are real and positive.

According to the above discussion, the neutrino mass and flavour bases are misaligned by the PMNS matrix
as follows,

\begin{align}
	\begin{aligned}
		\left(\begin{array}{c}
			\nu_{eL}  \\
			\nu_{\mu L} \\
			\nu_{\tau L} 
		\end{array}\right) = & \left(\begin{array}{ccc}
		U_{e1} & U_{e2} & U_{e3} \\
		U_{\mu 1} & U_{\mu 2} & U_{\mu 3} \\
		U_{\tau 1} & U_{\tau 2} & U_{\tau 3} 
		\end{array}\right) 
		 \left(\begin{array}{c}
			\nu_1  \\
			\nu_{2} \\
			\nu_{3} 
		\end{array}\right) \equiv U_{\text {PMNS }} \left(\begin{array}{c}
		\nu_1  \\
		\nu_{2} \\
		\nu_{3} 
		\end{array}\right),
	\end{aligned}
\end{align}
where $\nu_{eL}, \nu_{\mu L}, \nu_{\tau L}$ are the $SU(2)_L$ partners to the left-handed charged lepton mass eigenstates and
$\nu_{1,2,3}$ are the neutrinos in their mass basis, as in Fig.~\ref{angles}. Following the standard 
convention~\cite{ParticleDataGroup:2024cfk} we can describe the unitary matrix $U_{\text {PMNS }}$ in terms of three angles, one CP violation phase and two Majorana phases

\begin{align}
	\begin{aligned}
		U_{\text {PMNS }}= & \left(\begin{array}{ccc}
			1 & 0 & 0 \\
			0 & c_{23} & s_{23} \\
			0 & -s_{23} & c_{23}
		\end{array}\right)\left(\begin{array}{ccc}
			c_{13} & 0 & s_{13} e^{-\mathrm{i} \delta} \\
			0 & 1 & 0 \\
			-s_{13} e^{\mathrm{i} \delta} & 0 & c_{13}
		\end{array}\right)  \left(\begin{array}{ccc}
		c_{12} & s_{12} & 0 \\
		-s_{12} & c_{12} & 0 \\
		0 & 0 & 1
		\end{array}\right) P,
	\end{aligned}
\end{align}

\begin{align}
	= \left(\begin{array}{ccc}
		c_{12} c_{13} & s_{12} c_{13} & s_{13} e^{-i \delta} \\
		-s_{12} c_{23}-c_{12} s_{13} s_{23} e^{i \delta} & c_{12} c_{23}-s_{12} s_{13} s_{23} e^{i \delta} & c_{13} s_{23} \\
		s_{12} s_{23}-c_{12} s_{13} c_{23} e^{i \delta} & -c_{12} s_{23}-s_{12} s_{13} c_{23} e^{i \delta} & c_{13} c_{23}
	\end{array}\right) P,
	\label{eq:parametrisation}
\end{align}
where $P$ contains the Majorana phases

\begin{align}
   P= 	 \operatorname{diag}\left(1, e^{i \alpha_{21} / 2}, e^{i \alpha_{31} / 2}\right),
\end{align}
The current $3 \sigma$ parameters intervals coming from the global fit of the neutrino oscillation data
~\cite{Esteban:2020cvm,Capozzi:2021fjo,deSalas:2020pgw} 
are for the normal neutrino mass ordering, $m_1^2 < m_2^2 < m_3^2$,

\begin{align}
&	\theta_{12}=31^{\circ}-36^{\circ}, \quad   \theta_{23}=40^{\circ}-52^{\circ}, \quad  \theta_{13}=8.2^{\circ}-8.9^{\circ}, \quad  \\& \delta = 0^{\circ}-45^{\circ}\quad \& \quad 110^{\circ} -  360^{\circ}, \quad  \frac{\Delta_{21}^2}{10^{-5} \mathrm{eV}^2} = 6.8 - 8.0, \quad   \frac{\Delta_{31}^2}{10^{-3} \mathrm{eV}^2} = 2.4 - 2.6.
	\label{eq:exp-data}\quad \quad 
\end{align}
where $\Delta_{21}^2= m_2^2 - m_1^2$ and $\Delta_{31}^2= m_3^2 - m_1^2$. 
There is no measurement yet of the lightest physical neutrino mass $m_1$, which 
is not constrained by neutrino oscillations, and could take any value from zero up to 
the experimental and cosmological limit upper limits of around $0.1-0.2$ eV~\cite{Denton:2025jkt,eBOSS:2020yzd,Planck:2018vyg}.
For a normal neutrino mass {\it hierarchy}, $m_1\ll m_2\ll m_3$, 
\beq
m_2\approx \sqrt{\Delta_{21}^2}\approx 0.0086 \ {\rm eV}, \ \ \ \  m_3\approx \sqrt{\Delta_{31}^2}\approx 0.05 \ {\rm eV}.
\label{masshierarchy}
\eeq

Neutrinoless double beta decay experiments~\cite{KamLAND-Zen:2022tow,GERDA:2020xhi,CUORE:2019yfd,EXO-200:2019rkq,nEXO:2021ujk,LEGEND:2021bnm}
mentioned in the Introduction, are sensitive to the magnitude
of the first element of the Majorana neutrino mass matrix in the diagonal charged lepton basis $|m^{\nu }_{11}|\equiv m_{ee}$ corresponding to
\beq
m_{ee}\equiv \Big|\sum_{i=1}^3U_{ei}^2m_i\Big|, \label{mee}
\eeq
and will be able to fully probe the inverted neutrino mass squared region in the coming years.

The Tritium beta decay experiment KATRIN~\cite{KATRIN:2019yun,KATRIN:2021uub,Katrin:2024tvg}, also mentioned in the Introduction, is agnostic as to the nature of neutrino mass
(Dirac or Majorana) and 
is sensitive to the ``electron neutrino mass'' defined by
\beq
m_{\nu_e}^2\equiv \sum_{i=1}^3|U_{ei}|^2|m_i|^2. \label{mnue2}
\eeq
So far only limits on $m_{ee} < 0.1$ eV (depending on nuclear physics uncertainties) and $m_{\nu_e} < 0.45$ eV~\cite{Katrin:2024tvg} have been set, but in the future measurements could be made.

\section{Single Right-handed Neutrino (the simplest case)}
\label{SRHN}
\subsection{The Seesaw Mechanism}
In the SM the lepton doublets $L_{l}$ in Eq.~\ref{leptondoublets},
are split apart into left-handed charged leptons $l_{L}$ and left-handed 
neutrinos $\nu_{lL}$, where $l=e,\mu , \tau$. The charged leptons become massive but the neutrinos remain massless at the renormalisable level.
This is due to electroweak gauge invariance, and the fact that there are only Higgs doublets. 
Although they may receive mass from the non-renormalisable Weinberg operators in Eq.~\ref{Wop}, it would be preferable to have a renormalisable origin for neutrino mass, and one simple way to achieve this is by introducing right-handed chiral neutrinos $\nu_R$ which are electroweak 
singlets with zero hypercharge. The resulting seesaw mechanism~\cite{Minkowski:1977sc,Yanagida:1979as,GellMann:1980vs,Glashow:1979nm,Mohapatra:1979ia,Schechter:1980gr}\footnote{The original seesaw mechanism we shall describe is sometimes referred to as the type I or type Ia seesaw mechanism to distinguish it from the type Ib version discussed later, and other neutrino mass mechanisms which we shall not describe here, including type II~\cite{Magg:1980ut,Schechter:1980gr} with heavy triplet scalars and type III~\cite{Foot:1988aq,Ma:1998dn} with triplet fermions (for a survey see e.g. \cite{Ma:2009dk}).} 
provides a high energy completion of the non-renormalisable Weinberg operators in Eq.~\ref{Wop}.

To begin with let us introduce a single right-handed sterile neutrino $\nu_R$~\cite{King:1998jw,Davidson:1998bi} which is an electroweak singlet chiral field with zero hypercharge and electric charge. Crucially this means that its CP conjugate field\footnote{Here are elsewhere $\sigma_2$ is the second Pauli matrix.
The mass term can also be written as $M_{R}\nu_R^T \sigma_2 \nu_R $.} $\nu_R^c=i\sigma_2 \nu_R^*$ (a left-handed antineutrino field)
is also neutral under all gauge charges, which allows a new kind of mass term to be constructed, of the kind envisaged by E. Majorana in 1937, namely 
\begin{equation}
M_{R}\overline{\nu_R^c}\nu_R^{} .
\label{MRR}
\end{equation}
Clearly such a Majorana mass term would be forbidden if $\nu_R$ carried electric charge.
The size of the Majorana mass $M_R$ is arbitrary, and can range from zero to infinity, or at least the Planck mass.
It is also worth emphasising that the Majorana mass only involves two independent degrees of freedom, namely those of the Weyl spinor,
$\nu_R$, since its CP conjugate $\nu_R^c$ involves just the same degrees of freedom. This is unlike the Dirac mass term for the electron in Eq.\ref{Dirac} which couples two independent Weyl spinors $e_L$ and $e_R$ and therefore involves four independent degrees of freedom. However it is possible to describe the right-handed neutrino as a four component Majorana spinor,
\beq
N = 
\begin{pmatrix}
\nu_R^c  \\
\nu_R
\end{pmatrix}
\label{N}
\eeq
which is the analogue of the four component electron Dirac spinor in Eq.\ref{e}, but which in this case
only involves two independent degrees of freedom. In this notation the CP conjugate of the Majorana spinor in Eq.\ref{N} is equal to itself,
\beq
N^c=N
\eeq
The Majorana mass term then can be written
\beq
\frac{1}{2}M_R \overline{N}N = \frac{1}{2}M_R (\overline{\nu_R^c}\nu_R^{} + \overline{\nu_R}\nu_R^{c})
\label{M}
\eeq
It is common to refer to such a Majorana particle as being its own antiparticle since its four-component Majorana spinor is self-conjugate
\footnote
{This terminology makes sense since $\nu_R$ does not carry any quantum numbers.}. 
However the key thing to understand is that there are only two independent degrees of freedom describing Majorana particles such as the right-handed neutrino, which accounts for the factor of $1/2$ in Eq.\ref{M} as compared to the Dirac mass term for the electron in Eq.\ref{Dirac}.

\begin{figure}[t]
	\centering
	\includegraphics[width=.45\textwidth]{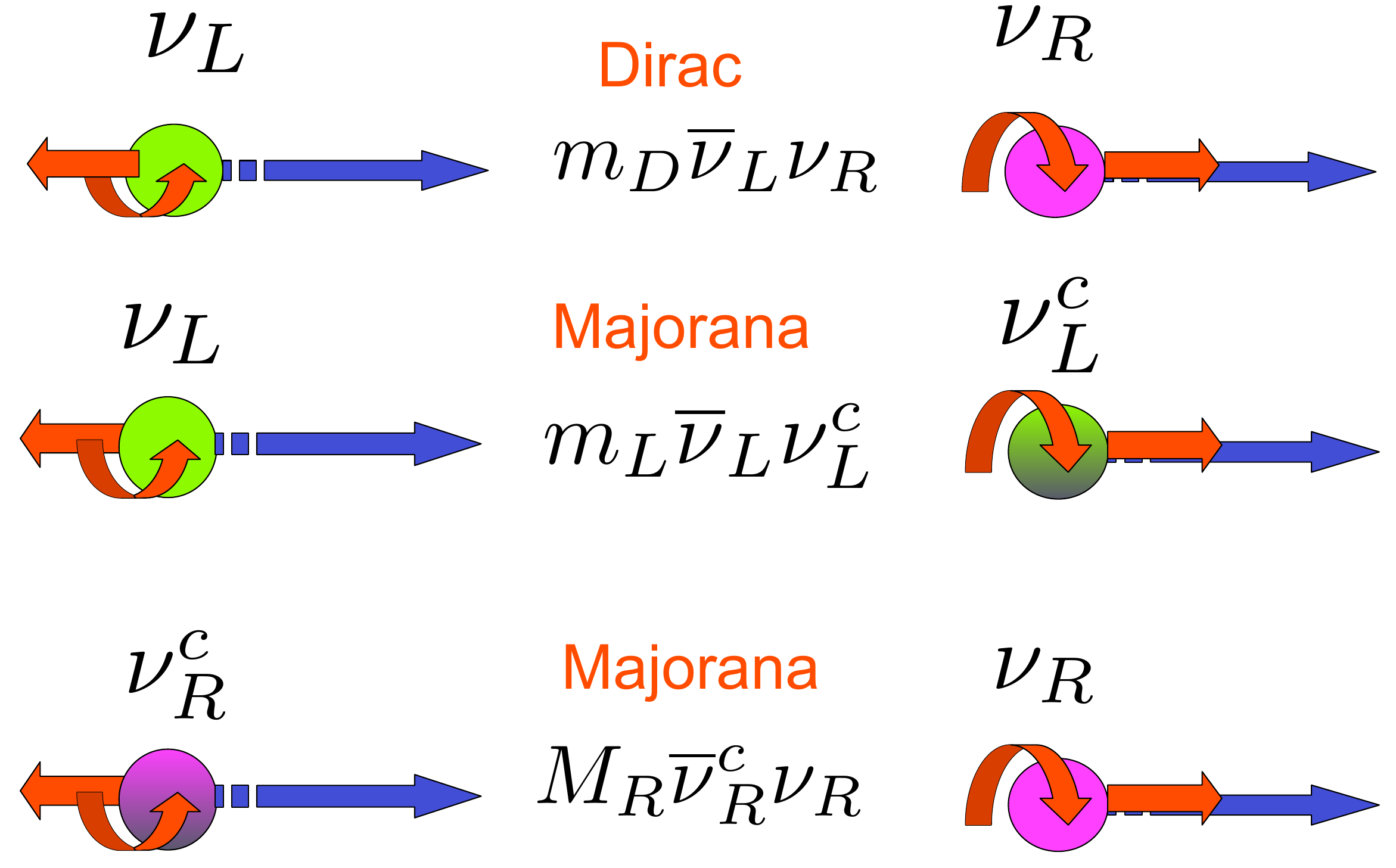}
	\caption{\footnotesize 
	Illustration of the different types of neutrino mass with one left-handed active neutrino $\nu_L$, and one right-handed sterile neutrino $\nu_R$. The long blue arrow is the momentum vector, while the short red arrow is the spin vector of the neutrino, where $L,R$ refers to left and right chirality, and $c$ indicates CP conjugation. The Dirac mass term involves both $L$ and $R$ degrees of freedom, while Majorana mass terms involve only either $L$ or $R$ degrees of freedom.}
	\label{numass}
\end{figure}

If we now recall the three massless chiral left-handed neutrinos $\nu_{eL}, \nu_{\mu L}, \nu_{\tau L}$ which remain below the electroweak symmetry breaking scale, further types of neutrino mass are possible.
As discussed below, it is possible to consider only one combination of these states $\nu_L$
which couples to the single right-handed neutrino $\nu_R$.
Ignoring electroweak interactions to begin with,
there are then three different types of mass term that can be written down: a Dirac mass $m_D$ connecting $\nu_L$ to $\nu_R$
\footnote{Dirac mass terms conserve lepton number.
Charged leptons also have this type of mass.}, a Majorana mass $m_L$ involving only the degrees of freedom of $\nu_L$, and a Majorana mass $M_R$ involving only the degrees of freedom of $\nu_R$,
as shown in Fig.~\ref{numass}.

In the Standard Model, the left-handed Majorana mass $m_L$ is forbidden by electroweak gauge invariance.
However Dirac masses $m_D$ may arise from Yukawa couplings 
\beq
y_{\nu} \overline{L} \tilde{H} \nu_R + H.c. \rightarrow y_{\nu}  \frac{v}{\sqrt{2}} (\overline{\nu}_L\nu_R + \overline{\nu}_R\nu_L)
\label{mD}
\eeq
involving $\tilde{H}=i\sigma_2H^*$, the same the Higgs doublet as
in Eq.~\ref{H} but with opposite hypercharge $Y=-1/2$,
where $y_{\nu}$ is a dimensionless coupling constant often called a neutrino Yukawa coupling.
We identify the Dirac mass for the neutrino as 
\beq
m_D= y_{\nu}  \frac{v}{\sqrt{2}}.
\label{mDD}
\eeq
Then including only $m_D$ and $M_R$ we can write the mass terms as a matrix,
\begin{equation}
\left(\begin{array}{cc} \overline{\nu_L} & \overline{\nu^c_R}
\end{array} \\ \right)
\left(\begin{array}{cc}
0 & m_{D}\\
m_{D} & M_{R} \\
\end{array}\right)
\left(\begin{array}{c} \nu_L^c \\ \nu_R \end{array} \\ \right)\,.
\label{matrix}
\end{equation}
Since the right-handed neutrinos are electroweak singlets the 
Majorana masses of right-handed neutrinos $M_{R}$
may be orders of magnitude larger than the electroweak
scale. Although the left-handed Majorana masses $m_L$ in the $11$ entry of the mass matrix 
are forbidden by electroweak symmetry
$SU(2)_L\times U(1)_Y$, they are generated effectively below the weak symmetry breaking scale, as follows.

The mass matrix in Eq.~\ref{matrix} may be diagonalised by,
\beq
\left(\begin{array}{cc}
\cos \theta & -\sin \theta\\
\sin \theta & \cos \theta \\
\end{array}\right)
\left(\begin{array}{cc}
0 & m_{D}\\
m_{D} & M_{R} \\
\end{array}\right)
\left(\begin{array}{cc}
\cos \theta & \sin \theta\\
-\sin \theta & \cos \theta \\
\end{array}\right)=
\left(\begin{array}{cc}
m_-& 0\\
0 & m_+\\
\end{array}\right)
\eeq
where
\beq
\tan 2\theta = \frac{2m_D}{M_R}.
\eeq
The mass eigenstates are, 
\beq
\left(\begin{array}{c}
\nu \\
N \\
\end{array}\right)
=
\left(\begin{array}{cc}
\cos \theta & -\sin \theta\\
\sin \theta & \cos \theta \\
\end{array}\right)
\left(\begin{array}{c} \nu_L^c \\ \nu_R \end{array} \\ \right)\,
\label{nuN}
\eeq
whose respective eigenvalues are,
\begin{equation}
m_{\mp}=\frac{M_R\mp \sqrt{M_R^2+4m_D^2}}{2}
\label{seesaw0}
\end{equation}
with trace $(m_+)+(m_-)=M_R$ and determinant $(m_+)(m_-)=-m_D^2$.
Thus, for fixed $m_D$, if one of the eigenvalues goes up the other goes down, like a seesaw.

\begin{figure}[t]
	\centering
	\includegraphics[width=.7\textwidth]{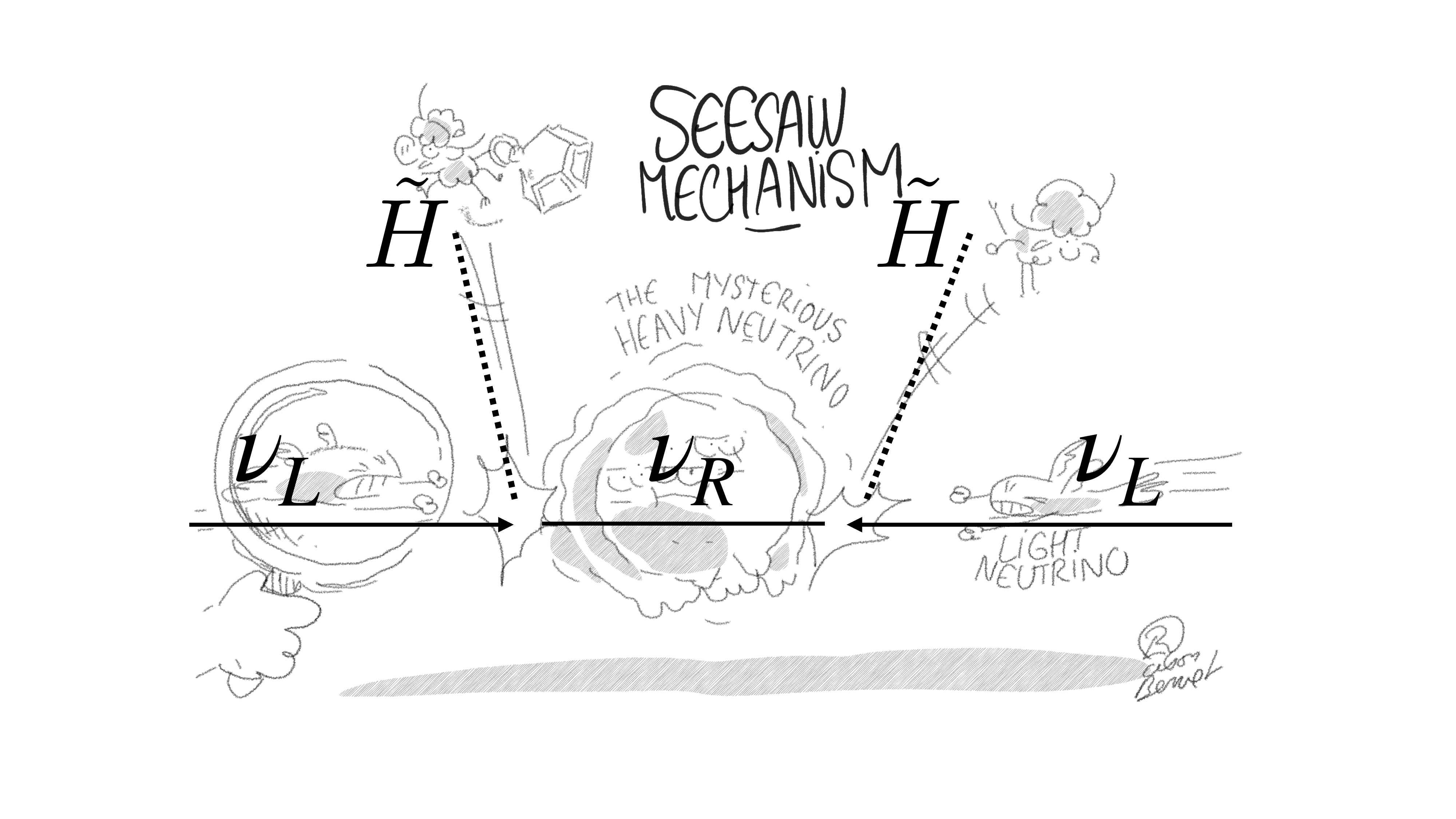}
	\caption{\footnotesize 
	The seesaw mechanism for one left-handed neutrino $\nu_L$ and one right-handed neutrino $\nu_R$. 
	Higgs doublets (the ``givers of mass'' depicted here carrying weights) allow the active looking ``light'' neutrinos $\nu_L$ 
	to couple to $\nu_R$ as in a Dirac mass term $m_D$. Gauge invariance allows an abitrarily large mass $M_R$ for the 
	(``mysterious heavy neutrino'') $\nu_R$. This results in a suppressed light effective left-handed Majorana mass 
	$m_{L}^{\rm eff}\sim m_D^2/M_R$ for $\nu_L$. The active light neutrinos $\nu_L$ are subject to direct experimental observation as indicated by the magnifying glass.}
	\label{typeIseesaw}
\end{figure}

\begin{figure}[t]
	\centering
	\includegraphics[width=.6\textwidth]{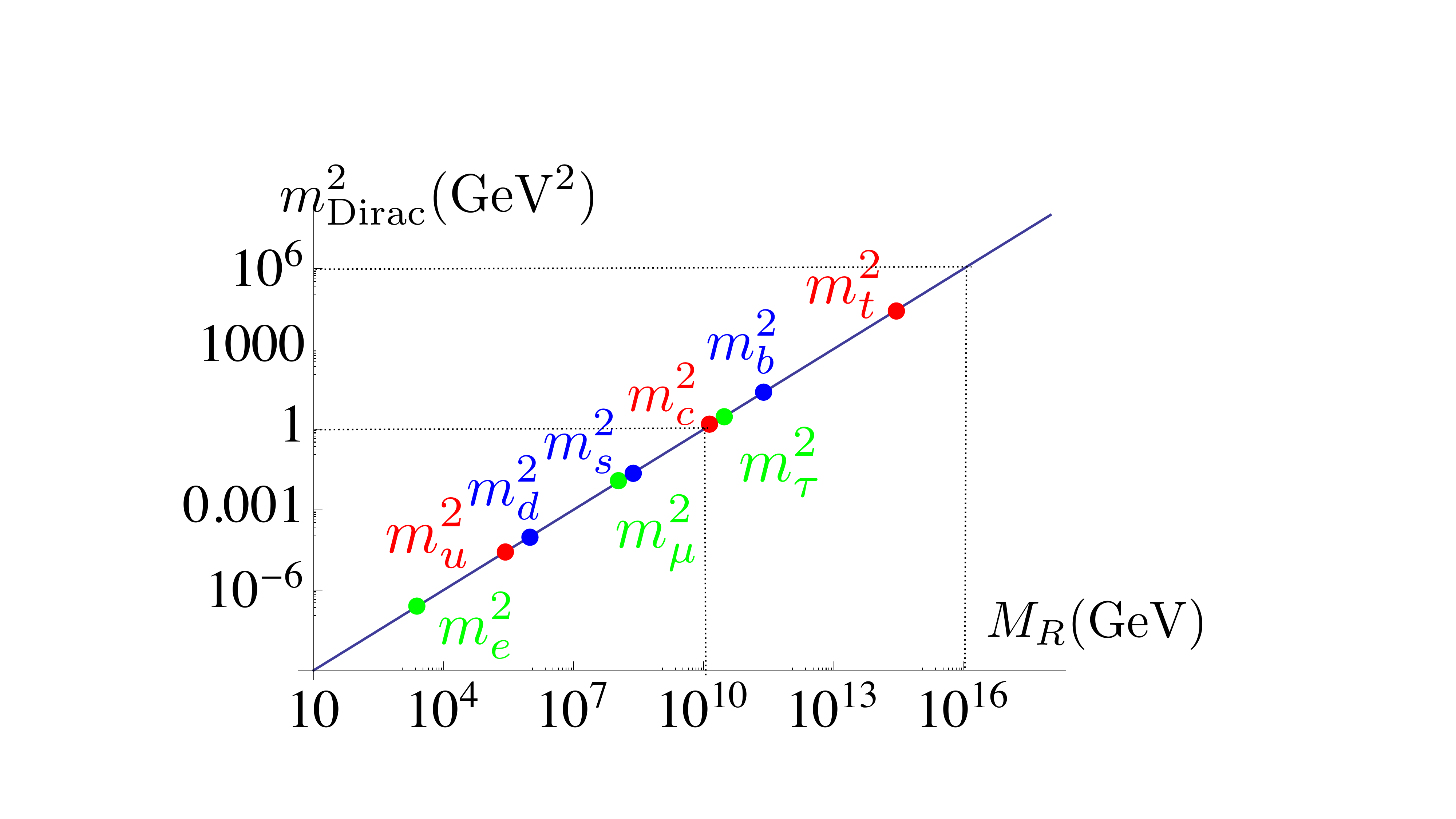}
	\caption{\footnotesize 
	Using the seesaw formula, $m_{L}^{\rm eff}\approx \frac{m_{D}^2}{M_{R}}\approx 0.1$ eV,
	we plot the square of the Dirac neutrino mass $m_D^2$ as a function of right-handed neutrino masses $M_R$. 
	For example the dotted lines show that $M_R=10^{16}$ GeV corresponds to $m_D=10^3$ GeV, while 
	$M_R=10^{10}$ GeV corresponds to $m_D=1$ GeV.
	 The squares of the charged quark and lepton masses are shown as coloured benchmark values of $m_D^2$.
	 For example taking $m^2_D=m^2_e$, the electron mass squared, would correspond to $M_R\approx 2.6$ TeV.}
	\label{seesawline}
\end{figure}

In the approximation that $M_{R}\gg m_{D}$
the matrix in Eq.~\ref{matrix} may be diagonalised by a small angle rotation with 
$\sin \theta \approx \tan \theta \approx \theta \approx m_D/M_R$,
and the heavier eigenvalue is approximately unchanged from the $22$ element, $m_+\approx M_R$, 
while the lighter eigenvalue $m_-$ fills in the zero element in the $11$ position and may thus be identified as a light effective left-handed Majorana neutrino mass,
\begin{equation}
m_-\approx m_{L}^{\rm eff}\approx -\frac{m_{D}^2}{M_{R}}\, ,
\label{seesaw}
\end{equation}
where the minus sign is practically irrelevant since it can be absorbed into the fermion field, and is often ignored.
This is the type I seesaw mechanism illustrated diagrammatically in Fig.~\ref{typeIseesaw}.
The key feature is that the effective left-handed Majorana mass $m_{L}^{\rm eff}$ is naturally
suppressed by the heavy mass $M_{R}$. Neither the Dirac neutrino mass $m_D$ nor the right-handed neutrino mass
$M_R$ are known, however experimentally we know that at least one of the light neutrino masses
must be around 0.1 eV.
Using this information in Fig.~\ref{seesawline} we use the seesaw formula in 
Eq.~\ref{seesaw} to plot possible values of $m_D$ and $M_R$ consistent with $m_{L}^{\rm eff}=0.1$ eV.

We can identify the left-handed neutrino state $\nu_L$ (above) as a linear combination of the 
coupling of the three left-handed neutrinos which couple to the single right-handed neutrino $\nu_R$~\cite{King:1998jw}, 
\beq
\overline{\nu}_R(d \nu_{eL}+ e\nu_{\mu L}+f\nu_{\tau L})\equiv m_D\overline{\nu}_R\nu_{L},
\label{SRHN1}
\eeq
where $m_D= \sqrt{|d|^2+|e|^2+|f|^2}$.
With the seesaw mechanism, it is thus possible to generate one light neutrino $\nu_{L}$ state with mass $m\approx m_D^2/M_R$,
which can be identified as the heaviest atmospheric neutrino, assuming a normal mass hierarchy~\cite{King:1998jw}.
Furthermore, if the couplings satisfy $d \ll e \sim f$, then we can generate an approximately maximal atmospheric
mixing angle and a small reactor angle~\cite{King:1998jw},
\beq
\tan \theta_{23}\approx \frac{|e|}{|f|} \sim 1, \ \ \ \  \theta_{13}\approx \frac{|d|}{\sqrt{|e|^2 + |f|^2}}.
\label{SRHN2}
\eeq

Thus the seesaw mechanism provides a high energy completion of the non-renormalisable Weinberg operators in Eq.~\ref{Wop}, with the right-handed neutrino mass $M_R$ providing an origin of the high energy mass scale $\Lambda$.
However, unlike the Weinberg operator, introducing right-handed neutrinos provides a testable prediction, namely the existence of heavy neutral leptons $N$, as we now discuss.

\subsection{The Heavy Neutral Lepton}

According to the above discussion, in the approximation that $M_{R}\gg m_{D}$,
the heavy neutral lepton mass eigenstate which we denote as $N$ in Eq.~\ref{nuN}
has a mass $m_+\approx M_R$ from Eq.~\ref{seesaw0}
and is dominantly identified as $\nu_R$ with a small admixture of $\nu_L^c$, while the light mass eigenstate which we denote as $\nu$ has mass $m_-\approx  m_{L}^{\rm eff}$ in Eq.~\ref{seesaw} is dominantly $\nu_L$ with a small admixture of $\nu_R^c$,\footnote{The approximation may be improved by expanding $\cos \theta \approx 1-\frac{1}{2}\theta^2$, leading to unitarity violation as discussed later.}
\beq
N \approx \nu_R + \theta \nu_L^c, \ \ \ \ 
\nu  \approx \nu_L -  \theta \nu_R^c, \ \ \ \ 
\label{admixtures}
\eeq
where 
\beq
\theta \approx  \frac{m_D}{M_R}, \ \ \ \  |\theta|^2 \approx  \frac{|m_{L}^{\rm eff}|}{M_R} \approx 10^{-10}\left( \frac{1 {\rm GeV}}{M_R}  \right),
\label{theta0}
\eeq
and the second equation follows from Eq.~\ref{seesaw} with $|m_{L}^{\rm eff}|\approx \frac{m_{D}^2}{M_{R}}\approx 0.1$ eV.
The heavy neutral lepton $N$ has suppressed but non-zero electroweak interactions since its mass eigenstate contains an admixture of the active neutrino $\nu_L^c$. This allows the heavy state $N$ to be produced experimentally, thereby providing a possible test of the seesaw mechanism.

To see this explicitly consider for example the left-handed couplings of the electron mass eigenstate $e_L$ to 
the heavy charged weak gauge boson $W^-$ (with similar results for the muon and tau),
\beq
-\frac{g}{\sqrt{2}}W_{\mu}^-\overline{e}_L\gamma^{\mu}{\nu}_{eL} + H.c.
\label{W}
\eeq
where $\nu_{eL}$ is the neutrino state which accompanies the electron mass eigenstate 
in the electroweak doublet in Eq.~\ref{leptondoublets}. Inverting Eq.~\ref{admixtures}, identifying $\nu_L$ with ${\nu}_{eL}$
\footnote{The analysis can be readily generalised to include ${\nu}_{eL}$, ${\nu}_{\mu L}$ and ${\nu}_{\tau L}$,
leading to similar results involving three mixing angles $\theta_{eN}\approx d/M_R$, $\theta_{\mu N}\approx e/M_R$,
$\theta_{\tau N}\approx f/M_R$, with the couplings given in Eq.~\ref{SRHN1}.},
\beq
\nu_R\approx N - \theta \nu^c, \ \ \ \ 
{\nu}_{eL}\approx \nu + \theta N^c, \ \ \ \ 
\label{admixtures2}
\eeq
which shows that ${\nu}_{eL}$ is approximately equal to the light mass eigenstate $\nu$ but contains a small admixture of the heavy neutral lepton $N^c$. From Eqs.~\ref{W}, \ref{admixtures2}, the $W$ coupling becomes, approximately,
\beq
-\frac{g}{\sqrt{2}}W_{\mu}^-\overline{e}_L\gamma^{\mu}
({\nu} + \theta N^c)
+ H.c.
\label{W2}
\eeq
where the first term in the bracket is approximately the usual coupling to the light neutrino state $\nu$, while the second term shows that there is a small coupling suppressed by the small angle $\theta$ (denoted $\theta_{eN}$ for the electron couplings) 
of the heavy neutral lepton $N^c$ to the heavy charged weak gauge boson $W^-$.
This, and similar terms involving the neutral weak gauge boson  $Z^0$ and the Higgs boson $h$, allows the heavy neutral lepton $N$ to be produced and decay in particle physics experiments, providing it is light enough to be produced. For example, using the seesaw formula in Eq.~\ref{seesaw} for a 
light neutrino $\nu$ with mass of 0.1 eV, taking $m^2_D=m^2_e$, the electron mass squared, would correspond to 
a heavy neutral lepton $N$ of mass $M\approx M_R\approx 2.6$ TeV, with the heavy-light mixing angle $\theta \approx m_D/M_R\approx 2\times 10^{-7}$.
Unfortunately, the heavy neutral lepton mass is too large and the heavy-light mixing angle too small for it to be discovered even at the LHC, 
where current LHC analyses lead to limits $\theta \lesssim 10^{-2}$ for $5 \ {\rm GeV} <  M < 50 \  {\rm GeV}$
~\cite{Drewes:2013gca,Drewes:2015vma,Drewes:2016jae}.

\begin{figure}[t]
	\centering
	\includegraphics[width=.6\textwidth]{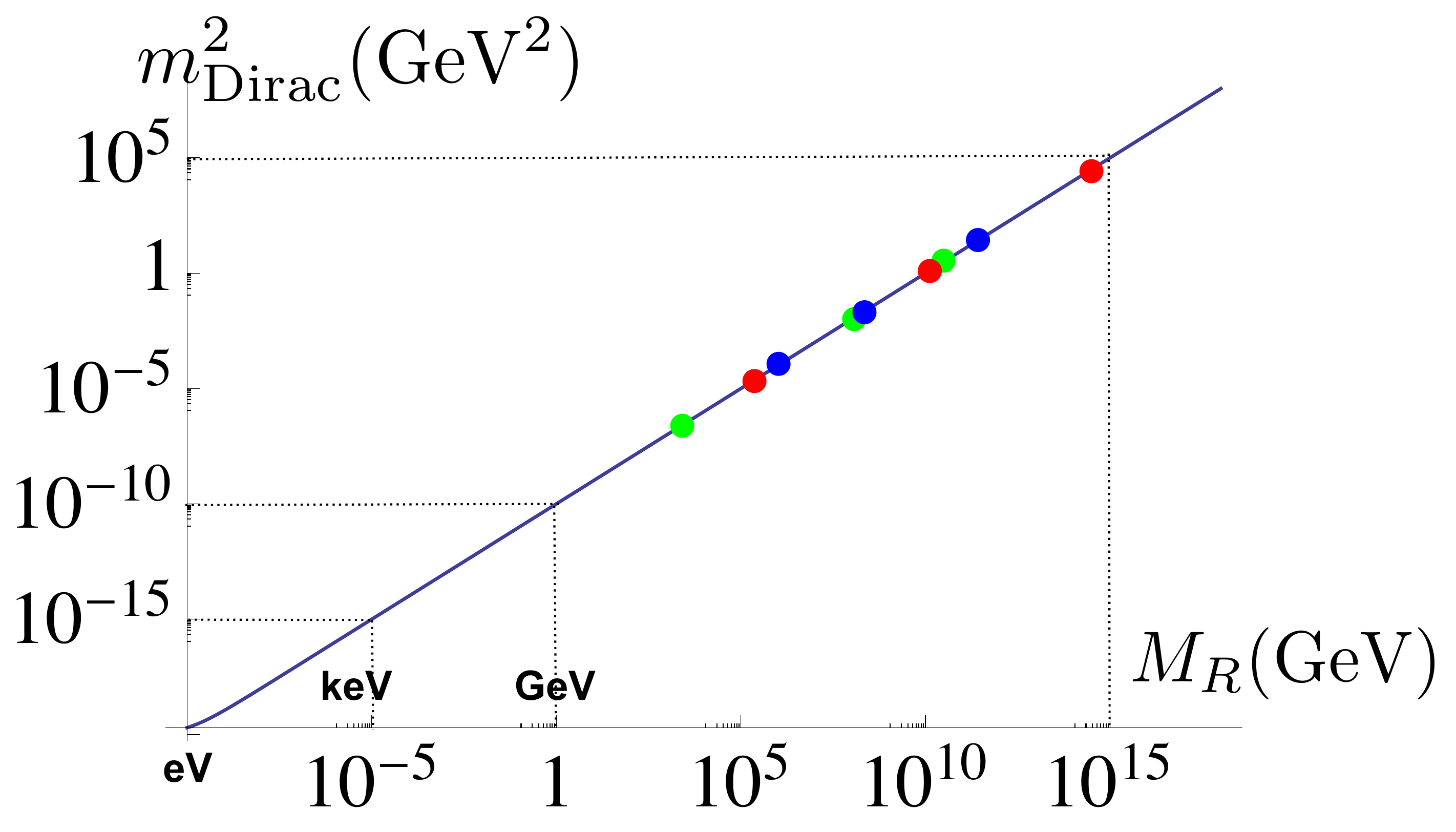}
	\caption{\footnotesize 
	The seesaw line, using the more accurate seesaw formula, $m_{L}^{\rm eff} \approx \frac{m_{D}^2 - (m_{L}^{\rm eff })^2}{M_{R}}$ for 
	$m_{L}^{\rm eff}\approx 0.1$ eV.
	We plot the square of the Dirac neutrino mass $m_D^2$ as a function of right-handed neutrino masses $M_R$
	down to 1 eV which may be accessible experimentally. The coloured dots show the charged fermion masses as in Fig.~\ref{seesawline}.
	}
	\label{steriles}
\end{figure}

It is possible that right-handed neutrino or sterile neutrino masses extend below the GeV scale, even though they can no longer be regarded as giving a completion to the Weinberg operator. The seesaw line may then be extended as in Fig.~\ref{steriles} for $M_R$ down to the GeV or keV scale where they may play the role of warm dark matter~\cite{Dolgov:2000ew,Abazajian:2001nj,Kusenko:2009up}. Such light right-handed neutrinos are then more often referred to as sterile neutrinos, as in for example the neutrino minimal SM \cite{Asaka:2005an,Asaka:2005pn,Boyarsky:2009ix} in which a keV sterile neutrino provides warm dark matter, while a pair of almost degenerate GeV scale sterile neutrinos can provide resonantly enhanced leptogenesis. If the right-handed neutrino masses are extended down to the eV scale, they affect neutrino oscillation physics, which takes us beyond the three neutrino oscillation paradigm considered here~\cite{Kopp:2011qd,Gariazzo:2015rra,Dentler:2018sju}.

\section{Three Right-Handed Neutrinos (the canonical case)}
\label{3RHN}

\subsection{Parametrising the Seesaw Mechanism}
Motivated by extensions of the SM in which baryon minus lepton number, $B-L$, symmetry is gauged, ranging from a simple 
$U(1)_{B-L}$ extension~\cite{King:2004cx}, to various levels of unification~\cite{Pati:1974yy,Fritzsch:1974nn,Senjanovic:1975rk},  
one right-handed neutrino per family of quarks and leptons is required to cancel gauge anomalies.
At low energies, below the $B-L$ symmetry breaking scale, the Lagrangian responsible for the lepton mass sector then consists of the SM plus three right-handed neutrinos with heavy Majorana mass,
\begin{equation}
{\cal L}^{\rm lepton}_{\rm mass}  = -H \overline{L}_i {y}^e_{ij} e_{Rj} 
-\tilde{H} \overline{L}_i {y}^{\nu}_{ia} \nu_{Ra} 
- \frac{1}{2}\overline{\nu_{Ra}^c}M_R^{ab} \nu_{Rb} +\text{H.c.}
 \ ,
\label{SM}
\end{equation}
in Weyl fermion notation in some arbitrary basis,
where SM family indices $i,j=1,2,3$ and right-handed neutrinos are labelled by $a,b=1,2,3$.
It is often convenient to work in the diagonal charged lepton and right-handed neutrino mass matrix basis, commonly referred to as the flavour basis, but above we work in an arbitrary basis where all matrices above are not diagonal in general. As in Eqs.~\ref{mD}, \ref{mDD} the Dirac mass matrix is generated after the Higgs VEV by
\beq
\tilde{H} \overline{L}_i {y}^{\nu}_{ia} \nu_{Ra}  \rightarrow \overline{\nu}_{Li} m^D_{ia}\nu_{Ra} 
\label{mD2}
\eeq
and, after integrating out the heavy right-handed neutrinos, the seesaw formula in Eq.~\ref{seesaw} is generalised to the matrix equation
\footnote{We have written $m^{\nu} = m_{L}^{\rm eff}$ as the light effective left-handed Majorana neutrino mass matrix, defined by,
$\frac{1}{2}\overline{\nu_{Li}}m^{\nu}_{ij} \nu^c_{Lj}$.}
\begin{equation}
m^{\nu} \approx -m_{D}M_R^{-1}m_{D}^T\,.
\label{seesaw2}
\end{equation}
Having integrated out the right-handed neutrinos, we are left with the lepton mass Lagrangian in Eq.~\ref{lepton}, 
where the neutrino mass matrix $m^{\nu}$ is given by Eq.~\ref{seesaw2}, 
leading to neutrino mass and mixing, as in the case of the Weinberg operator.

In the three right-handed neutrino seesaw mechanism, there are clearly more parameters than observables,
since the Dirac mass matrix $m_D$ is a $3\times 3$ matrix with arbitrary complex parameters, so this matrix alone has 18 free parameters. We now discuss a bottom-up approach to parameterising seesaw parameters, while constraining them to be consistent with neutrino oscillation data~\cite{Casas:2001sr}.

It is often convenient to work in the basis discussed in Eq.~\ref{diagchargedleptons1}, in which the charged lepton mass matrix is diagonal.
We also work in the diagonal right-handed neutrino mass basis, which together with the diagonal charged lepton mass basis, defines the 
flavour basis. In the flavour basis, the neutrino mass matrix $m^{\nu}$ is determined by the PMNS matrix,
using Eqs.~\ref{enu},\ref{diagonalisation},
\beq
m^{\nu}\approx U_{\rm PMNS}{\rm diag}(m_1, m_2, m_3)U^T_{\rm PMNS}.
\label{numassmatrix}
\eeq
The condition on the seesaw parameters for obtaining the correct neutrino observables can be obtained by combining 
Eqs.~\ref{seesaw2} and \ref{numassmatrix}, dropping the unphysical minus sign,
\beq
m_D {\rm diag}(M_1, M_2, M_3)^{-1}m_{D}^T= U_{\rm PMNS}{\rm diag}(m_1, m_2, m_3)U^T_{\rm PMNS}.
\label{condition}
\eeq
The choice of Dirac neutrino mass matrix $m_D$ consistent with Eq.~\ref{condition} is not unique.
It may be parameterised in a bottom-up way by taking the square root of Eq.~\ref{condition}, leading to,
\beq
m_D= U_{\rm PMNS}{\rm diag}(m_1, m_2, m_3)^{\frac{1}{2}}R^T{\rm diag}(M_1, M_2, M_3)^{\frac{1}{2}}
\label{CI}
\eeq
where $R$ is a complex orthogonal $3\times 3$ matrix which satisfies $R^TR= 1$ and
contains six real arbitrary parameters~\cite{Casas:2001sr}. Clearly there remains considerable freedom in the choice of parameters consistent with data.
The allowed parameter space has been very widely 
studied~\cite{Shaposhnikov:2008pf,Gavela:2009cd,Ruchayskiy:2011aa,Asaka:2011pb,Hernandez:2016kel,Caputo:2016ojx,Drewes:2016jae,Drewes:2018gkc,Das:2014jxa,Das:2015toa,Das:2016hof,Das:2018hph}, including analyses which 
specifically focus on leptogenesis~\cite{Buchmuller:2002rq,Buchmuller:2003gz,Buchmuller:2004nz,Blanchet:2006be,Davidson:2008bu},
including flavour dependent effects~\cite{Abada:2006fw,Nardi:2006fx,Abada:2006ea,Antusch:2010ms,DiBari:2012fz,DiBari:2021fhs,Drewes:2024pad}.

The seesaw mechanism provides a high energy completion of the non-renormalisable Weinberg operators in Eq.~\ref{Wop}, and comes with a prediction, namely the existence of right-handed (sterile) neutrinos or heavy neutral leptons $N_a$.  In what follows we shall extend the discussion of heavy neutral leptons
in Eqs.~\ref{admixtures}, \ref{admixtures2}, \ref{W} to include three families of leptons, including three right-handed neutrinos, and also discuss a new feature namely that, contrary to the case of the Weinberg operator, lepton mixing is no longer described by a unitary matrix.

\subsection{Non-Unitarity of the Lepton Mixing Matrix}

In the flavour basis the lepton mass Lagrangian after electroweak breaking becomes
\begin{equation}
{\cal L}^{\rm lepton}_{\rm mass}  = -\sum_{l=e,\mu , \tau} \overline{l}_L m_l l_{R} 
- \overline{\nu}_{lL}m^D_{la}\nu_{Ra}
- \frac{1}{2}\overline{\nu_{Ra}^c}M_R^{a} \nu_{Ra} +\text{H.c.}
 \ ,
\label{SM2}
\end{equation}
The full neutrino mass matrix is then as in Eq.~\ref{matrix} but is now a $6\times 6 $ matrix with the first three rows and columns corresponding to the active neutrinos are labeled by $l=e, \mu , \tau$ and the second three rows and columns by $a,b=1,2,3$, corresponding to the right-handed neutrinos. It can be block diagonalised by 
a $6\times 6 $ unitary matrix $U$~\cite{Schechter:1980gr},
\begin{equation}
U
\left(\begin{array}{cc}
0 & m_{D}\\
m_{D}^T & M_{R} \\
\end{array}\right) 
U^T
\approx 
\left(\begin{array}{cc}
m_{\nu} & 0 \\
0 & M_{R} \\
\end{array}\right) 
\label{matrix2}
\end{equation}
To leading order in small angles, $\theta_{la}$, which mix the right-handed sterile neutrino states with the left-handed active neutrino states
the $6\times 6 $ matrix may be written as~\cite{Ibarra:2011xn},
\beq
U \approx 
\begin{pmatrix}
1-\frac{1}{2}\theta \theta^{\dagger} & \theta \\
-\theta^{\dagger} & 1-\frac{1}{2}\theta^{\dagger}  \theta
\end{pmatrix}
\eeq
where 1 is the unit $3\times 3 $ matrix, and the matrix $\theta_{la}$ is given by the matrix form of Eq.~\ref{theta0}
\beq
\theta \approx  {m_D}M_R^{-1}.
\label{theta}
\eeq
The light neutrino mass matrix $m_{\nu}$ is given by the seesaw formula in Eq.~\ref{seesaw2},
\begin{equation}
m^{\nu} \approx -m_{D}M_R^{-1} m_{D}^T\approx -\theta m_{D}^T
\label{seesaw3}
\end{equation}
where $m_{\nu}$ is diagonalised by the unitary matrix $V_{\nu L}$ as in Eq.~\ref{diagonalisation},
\beq
V_{{\nu}_L}m^{\nu}V_{{\nu}_L}^{T}= {\rm diag} (m_1, m_2, m_3).
\label{diagonalisation}
\eeq
The three light neutrino mass eigenstates $(\nu_1, \nu_2, \nu_3)$ are related to the flavour eigenstates $\nu_l$ by,
\begin{align}
	\begin{aligned}
		\left(\begin{array}{c}
			\nu_{eL}  \\
			\nu_{\mu L} \\
			\nu_{\tau L} 
		\end{array}\right) 
		=
		(1-\frac{1}{2}\theta \theta^{\dagger})V^{\dagger}_{{\nu L}}
		 \left(\begin{array}{c}
		\nu_1  \\
		\nu_{2} \\
		\nu_{3} 
		\end{array}\right)
		\equiv 
		U_{\text {PMNS }}
		 \left(\begin{array}{c}
		\nu_1  \\
		\nu_{2} \\
		\nu_{3} 
		\end{array}\right),
	\end{aligned}
	\label{PMNS1}
\end{align}
so that, unlike Eq.~\ref{enu}, $U_{\text {PMNS}}$ as defined above is no longer exactly unitary. 

\subsection{Phenomenology of Heavy Neutral Leptons}

The couplings of $N_a$ to the charged weak gauge bosons $W^{\pm}$ in Eq.\ref{W} are then generalised to~\cite{Ibarra:2011xn},
\beq
-\frac{g}{\sqrt{2}}W_{\mu}^-\overline{l}_L\gamma^{\mu}
(U^{\rm PMNS}_{li}{\nu_i} + \theta_{la} N^c_a)
+ H.c.
\label{W2}
\eeq
where $l=e,\mu , \tau$ label the leptons (including neutrinos) in the charged lepton diagonal basis, 
while $i=1,2,3$ labels the three light neutrino mass eigenstates $\nu_i$, and 
$a=1,2,3$ labels the three neutral heavy lepton mass eigenstates $N_a$.
The heavy neutral leptons will also couple to the neutral weak gauge boson $Z^0$ and the Higgs boson $h$ through the small
heavy-light mixing angles $\theta$, leading to interesting collider phenomenology~\cite{delAguila:2008cj,Atre:2009rg,Dev:2013wba,Cai:2017mow,Pascoli:2018heg, Ibarra:2011xn,Chakraborty:2018khw,Curtin:2018mvb,Abdullahi:2022jlv}.

\begin{figure}[t]
	\centering
	\includegraphics[width=.5\textwidth]{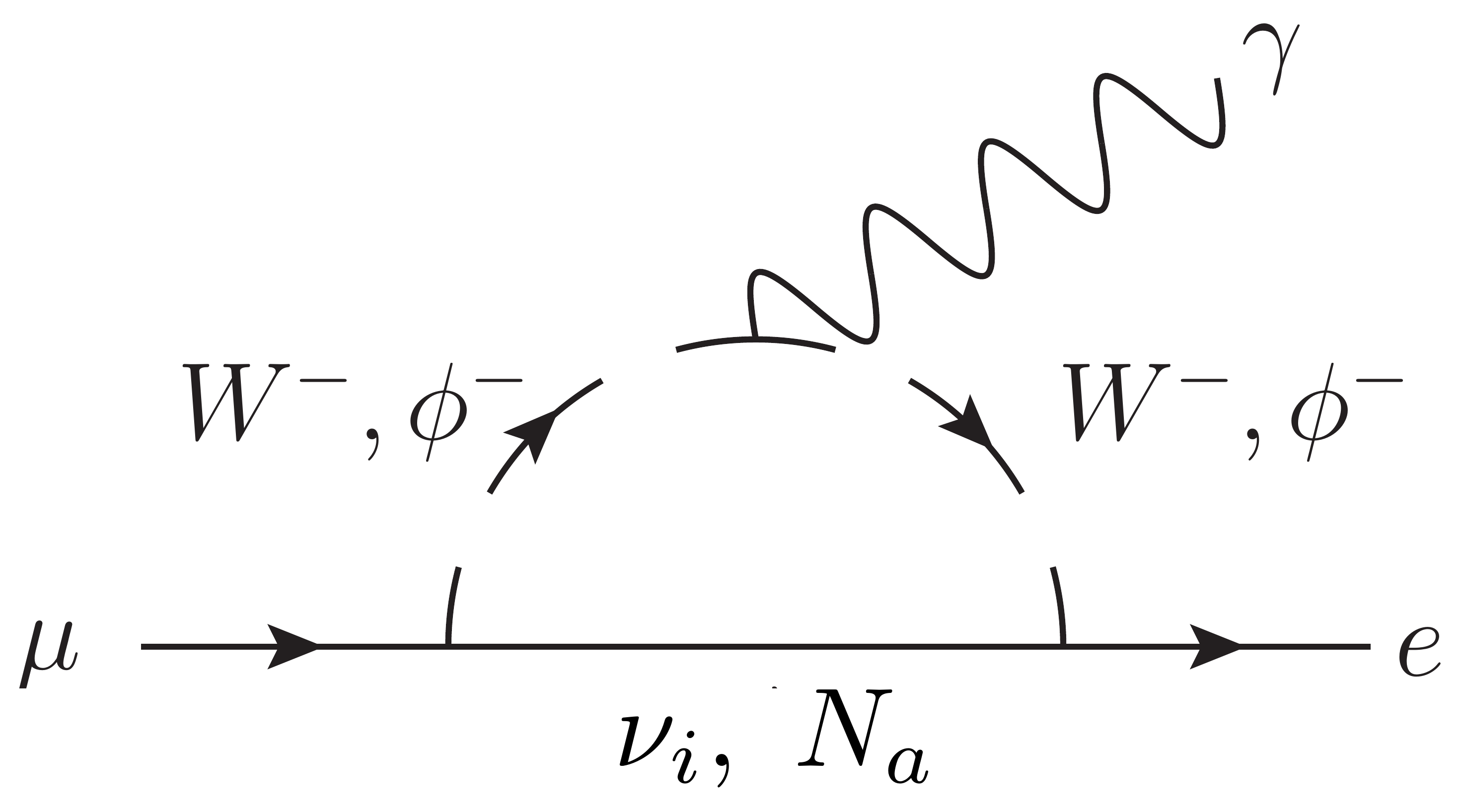}
		\caption{\footnotesize     
	The non-unitarity of the PMNS matrix can lead to enhanced rates for $\mu\rightarrow e \gamma$ via the exchange of virtual light and heavy neutrino mass eigenstates in the one loop diagram above. The couplings are given in Eq.~\ref{W2} and $\phi^-$
	is the Goldstone boson component of $W^-$ according to the Higgs mechanism. 
	Figure is adapted from \cite{Hernandez-Garcia:2019uof}.
		}
	\label{meg}
\end{figure}

The non-unitarity of the PMNS matrix in Eq.~\ref{PMNS1}, with the $W$ couplings in Eq.~\ref{W2},
leads to enhanced lepton flavour violation
processes such as $\mu \rightarrow e \gamma$ as shown in Fig.~\ref{meg}~\cite{Antusch:2006vwa,Gavela:2009cd,Hernandez-Garcia:2019uof}. This is one testable difference between just having the Weinberg operator and having the seesaw mechanism with right-handed neutrinos. Another testable difference is the prediction of heavy neutral leptons, i.e. the heavy mass eigenstates $N_a$ with masses $M_{Ra}$, which can be produced in experiments providing they are not too heavy and their heavy-light mixing angles are not too small.

\begin{figure}[t]
	\centering
	\includegraphics[width=.45\textwidth]{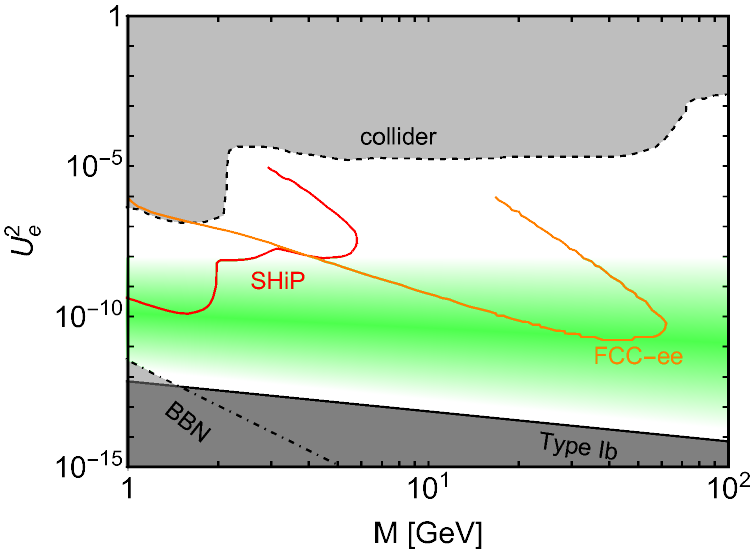}
	\includegraphics[width=.45\textwidth]{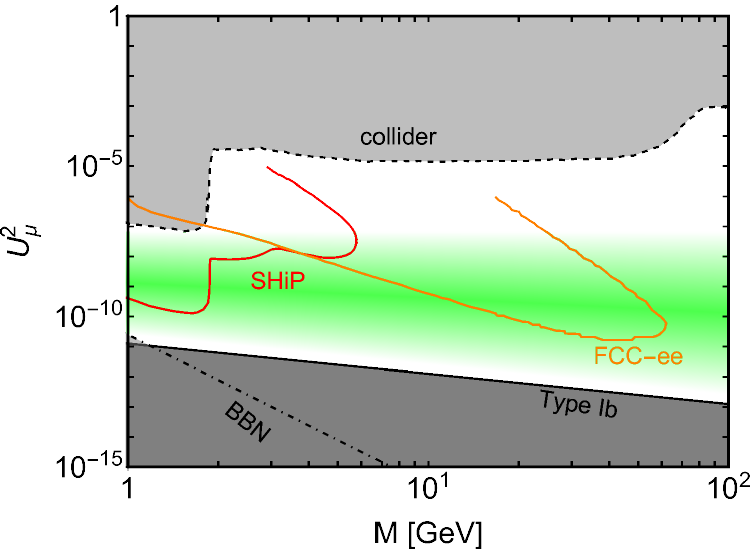}
	\caption{\footnotesize 
	This shows present bounds and future prospects on measuring the neutrino-sterile mixing parameter, for 
		$\theta_{eN}^2 \sim U_{e}^2$
		and $\theta_{\mu N}^2 \sim U_{\mu }^2$, as a function of the heavy neutral lepton mass $M$. 
	The whole white and green parameter space is allowed, with the green parameter space allowed by leptogenesis in some models~\cite{Fu:2021fyk}. The existing collider data~\cite{Drewes:2013gca,Drewes:2015vma,Drewes:2016jae} constrants are shown, and future experiment reach from SHiP~\cite{SHiP:2018xqw} and FCC-$ee$~\cite{Blondel:2014bra}. 
The lower bounds coming from the seesaw mechanism and Big Bang Nucleosynthesis (BBN) (due to the late decaying HNL) are also shown (from~\cite{Fu:2021fyk}). It is worth mentioning that the green parameter space allowed by leptogenesis can be further extended down to the seesaw line and up to to the estimated sensitivities of the LHC main detectors in some scenarios~\cite{Drewes:2021nqr}. }
	\label{U2}
\end{figure}

\begin{figure}[t]
	\centering
	\includegraphics[width=.4\textwidth]{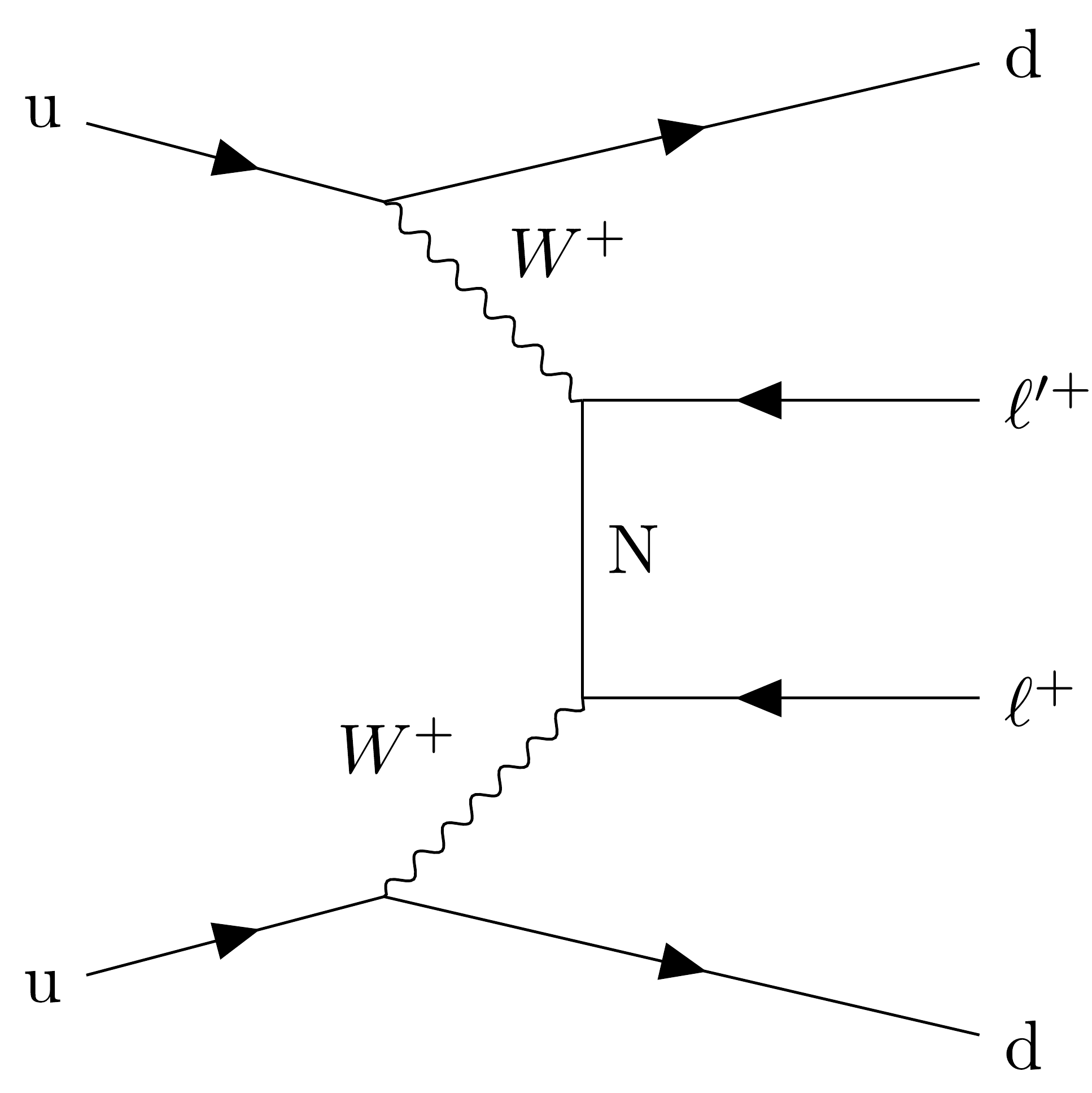}
	\caption{\footnotesize 
A Feynman diagram contributing to the possible indirect detection of HNLs at the LHC via $W^+W^+$ scattering, leading to mass limits in the range 50 GeV to 20 TeV~\cite{CMS:2022hvh,ATLAS:2024rzi}. 
The same sign dileptons in the final state 
are indicative of lepton number violation due to the Majorana nature of the intermediate HNL state $N$. The diagram involving $W^-W^-$ scattering is similar to that for neutrinoless double beta decay in Fig.~\ref{ndbc}.}
	\label{WW}
\end{figure}

\begin{figure}[t]
	\centering
	\includegraphics[width=.7\textwidth]{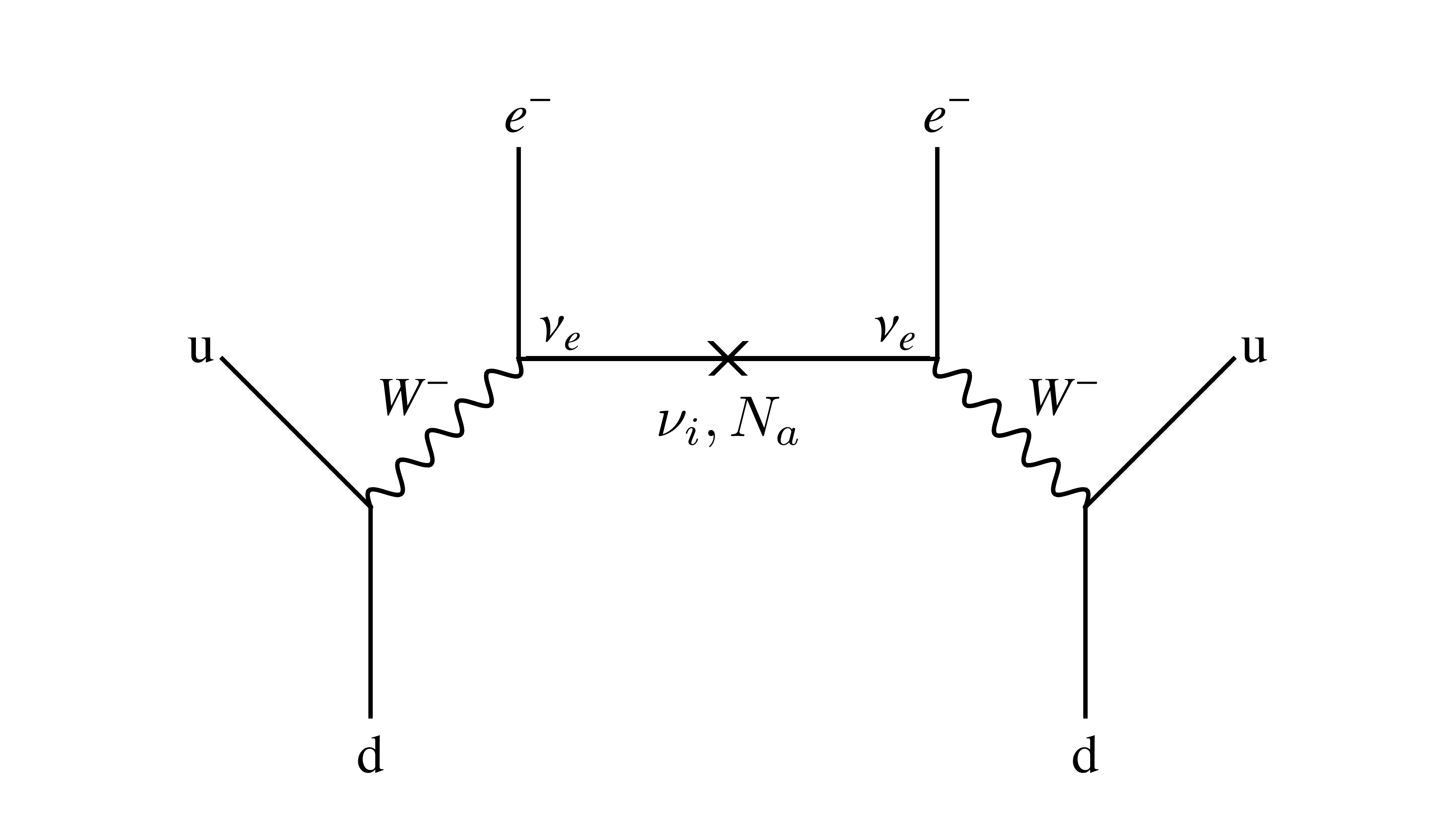}
	\caption{\footnotesize 
	The diagram which contributes to the amplitude for neutrinoless double beta decay involves the annihilation of the electron neutrino.
	To calculate the amplitude, the electron neutrino state $\nu_e$ must be expanded in terms of the mass eigenstates $\nu_i$ and $N_a$, which propagate in the diagram. }
	\label{ndbc}
\end{figure}

In the small angle approximation assumed here $U_{la} \approx   \theta_{la} $ and experimental limits usually are 
set on $|U_{lN}|^2\sim U_l^2$ for  $l=e,\mu , \tau$ as a function of some generic heavy neutral lepton $N$ of mass $M$,
as shown in Fig.~\ref{U2}.
For masses $M$ below the TeV scale, the heavy-light mixing is constrained by existing collider data~\cite{Drewes:2013gca,Drewes:2015vma,Drewes:2016jae} and future experiments~\cite{Deppisch:2015qwa,Chianese:2018agp,Beacham:2019nyx} for example, the SHiP experiment~\cite{SHiP:2018xqw} and FCC-$ee$~\cite{Blondel:2014bra}. 
For instance, LHC searches have been performed for $N$ produced and decaying through its $W$ couplings, for example direct production of HNLs through $W$ production and decay,
$W\rightarrow e + N$ production and $N\rightarrow  e  + W$, with the signature of two isolated charged leptons plus a $W$.
Current LHC analyses lead to limits $|U_{eN}|^2\lesssim 10^{-5}$ for $5 \ {\rm GeV} <  M < 50 \  {\rm GeV}$. Indirect detection of HNLs also provides important limits, for example via $WW$ scattering, as in 
Fig.~\ref{WW}, where the Majorana nature of the intermediate $N$ results in same sign dileptons in the final state (the collider equivalent of neutrinoless double beta decay),
leading to limits in the mass range 50 GeV to 25 TeV~\cite{CMS:2022hvh,ATLAS:2024rzi}.

The question then arises of whether it is possible to obtain such sizeable 
mixing angles for heavy neutral leptons with masses accessible at colliders,
consistent with the seesaw formula for the neutrino mass matrix which satisfies the neutrino oscillation data. The answer is yes in general, since there is considerable freedom in the choice of seesaw parameters consistent with neutrino phenomenology, namely the three right-handed neutrino masses $M_1, M_2, M_3$ and the choice of Dirac mass matrix $m^D$ which is parameterised by the six real arbitrary parameters of the complex orthogonal matrix $R$ in Eq.~\ref{CI}, so that Eq.~\ref{theta},
gives,
\beq
\theta \approx  U_{\rm PMNS}{\rm diag}(m_1, m_2, m_3)^{\frac{1}{2}}R^T{\rm diag}(M_1, M_2, M_3)^{-\frac{1}{2}}.
\label{CI2}
\eeq
If the complex orthogonal matrix $R$ contains large elements then the mixing angles may be enhanced compared to the one right-handed neutrino estimate in Eq.~\ref{theta1} of $\theta\sim {m^{\frac{1}{2}}} {M^{-\frac{1}{2}}}$.

Neutrinoless double beta decay in Fig.~\ref{ndbc} involves the electron neutrino $\nu_e$, expanded in terms of the light mass eigenstates 
$\nu_i$ and heavy mass eigenstates $N_a$, which couple to the $W$ as shown in Eq.~\ref{W2} for $l=e$.
The nuclear half-life depends on the effective neutrinoless double beta decay parameter $m_{\beta \beta}$, given by,
\beq
m_{\beta \beta} =\Big| \sum_{i=1}^3{U^{\rm PMNS}_{ei}}^2m_i + \sum_{a=1}^3\theta^2_{ea}M_a\frac{{\cal A}(M_a)}{{\cal A}(0)}\Big| \label{mee2},
\eeq
where the first sum is approximately equal to $m_{ee}$ in Eq.~\ref{mee}, while the second sum involves the heavy neutral leptons
and depends on the mass dependent hadronic amplitudes ${\cal A}(M_a)$, which capture all the hadronic and nuclear physics. 
It is commonly assumed that~\cite{Blennow:2010th,Mitra:2011qr,Li:2011ss},
\beq
\frac{{\cal A}(M_a)}{{\cal A}(0)}\approx \frac{p^2}{p^2+M_a^2}, 
\eeq
where the typical hadronic physics energy 
$p\sim 100$~MeV (of order the pion mass), although a full effective field theory approach is preferred~\cite{Dekens:2024hlz}.
For $M_a^2\ll p^2$, the hadronic amplitude ratio becomes unity and the two sums can be combined to give a result
equivalent to the first entry of the full $6\times 6 $ neutrino mass matrix, which is zero (see left-hand side of Eq.~\ref{matrix2}). 
For $M_a^2\gtrsim p^2$, neutrinoless double beta decay will be non-zero in the seesaw mechanism, 
with an interesting interplay between neutrinoless double beta decay and leptogenesis~\cite{deVries:2024rfh}.
For $M_a^2\gtrsim $ a few GeV, the result is well approximated by the first sum in Eq.~\ref{mee2}, with the other terms heavily
suppressed by factors of $\frac{p^2}{M_a^2}$.

\section{Two Right-Handed Neutrinos (the minimal case)}
\label{2RHN}

\subsection{Parametrisation}
Although the seesaw mechanism can qualitatively explain the smallness of neutrino masses through the heavy right-handed neutrinos (RHNs), if one doesn't make other assumptions, it contains too many parameters to make any particular predictions for neutrino mass and mixing. 
In this section we consider the minimal case, consistent with all current neutrino data,
of a two right-handed neutrino (2RHN) model ~\cite{King:1999mb,King:2002nf,Frampton:2002qc} where the lightest neutrino is massless $m_1=0$. This can be regarded as the limiting case of three right-handed neutrinos, where one of the right-handed neutrinos has a negligible contribution to the seesaw mechanism, and so can be regarded as being decoupled. 

The Lagrangian is as in Eq.~\ref{SM} where SM family indices are $i,j=1,2,3$ as before but now there are only two right-handed neutrinos labelled by $a,b=1,2$. In the flavour basis, the Lagrangian below the electroweak scale becomes as in Eq.\ref{SM2}, 
\beq
{\cal L}^{\rm lepton}_{\rm mass}  = -\sum_{l=e,\mu , \tau} \overline{l}_L m_l l_{R} 
- \overline{\nu}_{lL}m^D_{l1}\nu_{R1}
- \overline{\nu}_{lL}m^D_{l2}\nu_{R2}
- \frac{1}{2}\overline{\nu_{R1}^c}M_1 \nu_{R1}
- \frac{1}{2}\overline{\nu_{R2}^c}M_2 \nu_{R2} 
 +\text{H.c.}
\eeq
The Dirac neutrino mass matrix with two right-handed neutrinos is a $3\times 2$ 
matrix with two columns $m^D_{l1}$ and $m^D_{l2}$, where $l=e, \mu , \tau$, which
can be parameterised by 
one complex angle $z$ or two real arbitrary parameters \cite{Ibarra:2003up},
\begin{eqnarray}
m^D_{l1} & = &  \sqrt{M_1}\left( \sqrt{m_2}\cos z U^{\rm PMNS}_{i2} + \sqrt{m_3}\sin z U^{\rm PMNS}_{i3} \right) \nonumber \\
m^D_{l2} & = &  \sqrt{M_2}\left(\sqrt{m_2}\sin z U^{\rm PMNS}_{i2} - \sqrt{m_3}\cos z U^{\rm PMNS}_{i3} \right),
\label{CI2}
\end{eqnarray}
for the case of normal neutrino mass squared ordering.
Towards the goal of predictivity, the two right-handed neutrino model is clearly a step forwards, since $m_D$ only involves
linear combinations of two columns of the PMNS matrix, parameterised by a single complex angle $z$,
in this bottom-up approach, rather than three,
as in Eq.~\ref{CI}, parameterised by three complex angles.
The general two right-handed neutrino model is indeed quite testable~\cite{Hernandez:2016kel,Drewes:2016jae,Drewes:2018gkc}, and leads to a quite restrictive parameter space when leptogenesis is considered~\cite{Asaka:2005pn,Shaposhnikov:2008pf,Canetti:2010aw,Asaka:2011wq,Antusch:2010ms,Antusch:2011nz,Canetti:2012vf,Canetti:2012kh,Shuve:2014zua,Abada:2015rta,Hernandez:2015wna,Drewes:2016lqo,Drewes:2016jae,Hernandez:2016kel,Asaka:2016zib,Drewes:2016gmt,Asaka:2017rdj,Abada:2017ieq,Antusch:2017pkq}.

For example, consider the case of 
Eq.~\ref{CI2} with $\sin z\approx 1$ and $\cos z\approx 0$,
\begin{eqnarray}
m^D_{l1} & \approx &  \sqrt{M_1}\sqrt{m_3} U^{\rm PMNS}_{l3}  \nonumber \\
m^D_{l2} & \approx &  \sqrt{M_2}\sqrt{m_2} U^{\rm PMNS}_{l2}.
\label{CI3}
\end{eqnarray}
In this case the first column of $m_D$, corresponding to the couplings of the first right-handed neutrino, are dominantly responsible for the heaviest physical neutrino mass $m_3$ and atmospheric mixing angle $\theta_{23}$, while the second column of $m_D$, corresponding to the couplings of the subdominant second right-handed neutrino, are responsible for the physical neutrino mass $m_2$ and the solar mixing angle $\theta_{12}$. 

However ultimately this reverse engineering approach does not tell us anything about the physics responsible for 
$m_D$ in the first place, although it is useful for classifying equivalent seesaw models \cite{King:2006hn}.
For example, if some model leads to a ``texture'' zero element of $m_D$ enforced by some symmetry, this will fix the angle
$z$ to some precise value, which looks like a fine tuning, while in fact it might be a natural consequence of the theory.
In pursuit of predictive models, we shall now abandon the bottom-up approach, and focus on cases
where $m_D$ is fixed by some top-down theory.

\subsection{Sequential Dominance of Three Right-Handed Neutrinos}

In this subsection we consider the conditions which can naturally lead to a neutrino mass hierarchy 
\beq
m_1\ll m_2\ll m_3,
\eeq
together with large neutrino mixing angles, without any tuning or cancellations of parameters.
Such cancellations can be avoided if each column of the Dirac mass matrix is associated mainly with a particular physical neutrino mass, an approach known as 
sequential dominance (SD)~\cite{King:1998jw,King:1999cm,King:1999mb,King:2002nf} of right-handed neutrinos.
Historically, SD was proposed before the general parameterisation in the previous subsection.
While the general parameterisation is a bottom-up model independent approach, SD is a top-down model dependent approach,
which can lead to predictions for the neutrino observables.

The basic idea of SD is that, in the flavour basis
(diagonal RHNs and charged lepton masses), one of the RHNs $\nu_R^{\mathrm{atm}}$ with mass 
$M_{\mathrm{atm}}$ is dominantly responsible for the heaviest physical neutrino mass $m_3$, while 
a second subdominant RHN $\nu_R^{\mathrm{sol}}$ with mass $M_{\mathrm{sol}}$ is mainly responsible for the second heaviest physical mass $m_2$, and a third essentially decoupled RHN $\nu_R^{\mathrm{dec}}$ of mass $M_{\mathrm{dec}}$ gives a very suppressed lightest neutrino mass $m_1$. This is the scenario anticipated in the notation of Fig.~\ref{blocks}, based on a limiting case of three right-handed neutrinos as predicted for example by $SO(10)$ GUTs, where particular models can lead to a strongly hierarchical and diagonal right-handed neutrino mass matrix \cite{King:2001uz,King:2003rf}.
In the limit that the third right-handed neutrino responsible for the lightest light neutrino mass is decoupled from the seesaw mechanism,
this leads to an effective 2RHN model with a neutrino mass hierarchy,
with $m_1 \approx 0$,
where the large neutrino mixing angles arise in a natural way from ratios of couplings to the same right-handed neutrino~\cite{King:1998jw,King:1999cm,King:1999mb,King:2002nf}, as we now discuss.

The two right-handed neutrinos have the following diagonal heavy Majorana mass matrix, in a notation
which is agnostic as to their mass ordering,
\begin{align}
	M_R=\left(\begin{array}{cc}
		M_{\mathrm{atm}} & 0 \\
		0 & M_{\mathrm{sol}}
	\end{array}\right).	
	\label{eq:RH-masse}
\end{align}
The Dirac neutrino mass matrix in the flavour basis is written in a simple notation as,
\begin{align}
	m^D=\left(\begin{array}{cc}
		d & a \\
		e & b \\
		f & c
	\end{array}\right), 
	\label{3by2}
\end{align}
where the first (second) column contains the couplings 
to the atmospheric (solar) RH neutrino, 
\beq
\nu_R^{\mathrm{atm}}(d \nu_{eL}+ e\nu_{\mu L}+f\nu_{\tau L}) +
\nu_R^{\mathrm{sol}}(a \nu_{eL}+ b\nu_{\mu L}+c\nu_{\tau L}).
\label{schematic}
\eeq
The atmospheric neutrino couplings will dominate the seesaw mechanism as in the single right-handed neutrino case
in Eq.~\ref{SRHN1}. 
\footnote{It is instructive to compare Eq.~\ref{CI3} to Eq.~\ref{3by2} with the condition in Eq.~\ref{SD1}. In both cases the first column of the Dirac mass matrix
dominates and leads to a natural neutrino mass hierarchy. However Eq.~\ref{CI3} imposes the stronger requirement that the first 
(second) column of the Dirac matrix is proportional to the third (second) column of the PMNS matrix.
On the other hand, Eq.~\ref{3by2} makes no such requirement and hence is more general.
For example, Eq.~\ref{3by2} allows for the possibility of a texture zero which would violate Eq.~\ref{CI3}.
The special case of Eq.~\ref{3by2} where Eq.~\ref{CI3} is satisfied is known as 
form dominance~\cite{King:2006hn,King:2009qt,Chen:2009um,Choubey:2010vs,King:2010bk,King:2011ab,King:2012vj}.}

Using the seesaw formula in Eq.~\ref{seesaw2}, with the matrices in Eqs.~\ref{eq:RH-masse}, \ref{3by2}, 
dropping the overall physically irrelevant minus sign,
the light effective left-handed Majorana neutrino mass matrix
$m^{\nu}$ can be written as,
\begin{equation}
m^{\nu}=
\frac{1}{M_{\rm atm}}\left( \begin{array}{ccc}
d^2 & de & df  \\
de & e^2 & ef\\
df & ef & f^2
\end{array}
\right) +
\frac{1}{M_{\rm sol}}\left( \begin{array}{ccc}
a^2 & ab & ac  \\
ab & b^2 & bc \\
ac & bc & c^2
\end{array}
\right).
\label{2rhn}
\end{equation}
The SD conditions are that $\nu_R^{\mathrm{atm}}$ (the first matrix above) dominates the seesaw mechanism,
\begin{align}
	  \frac{(d,e, f)^2}{M_{\mathrm{atm}}} \gg \frac{(a, b, c)^2}{M_{\mathrm{sol}}}.
	 \label{SD1}
\end{align}
Ignoring phases\footnote{For the full results including phases see~\cite{King:2002nf}.}, this leads to the approximate results for the neutrino parameters in the flavour basis, assuming $d \ll e, f$,
\begin{equation}
\tan\theta_{23}\approx\frac{|e|}{|f|}
\label{theta23}
\end{equation}
\begin{equation}
\tan\theta_{12}\approx \frac{a}{\cos\theta_{23}b-\sin\theta_{23}c}
\label{theta12}
\end{equation}
\begin{equation}
\theta_{13}\approx\frac{1}{m_{3}}\left[\frac{a\left(\sin\theta_{23}b+\cos\theta_{23}c\right)}{M_{\mathrm{sol}}}+\frac{d\sqrt{|e|^{2}+|f|^{2}}}{M_{\mathrm{atm}}}\right]
\end{equation}
\begin{equation}
m_{3}\approx\frac{|e|^{2}+|f|^{2}}{M_{\mathrm{atm}}}
\label{m3}
\end{equation}
\begin{equation}
m_{2}\approx \frac{a^{2}}{M_{\mathrm{sol}}}+\frac{(\cos\theta_{23}b-\sin\theta_{23}c)^{2}}{M_{\mathrm{sol}}}, \ \ \ \ 
m_1\approx 0.
\label{m2}
\end{equation}
These results show that the condition in Eq.~\ref{SD1} that $\nu_R^{\mathrm{atm}}$ dominates the seesaw mechanism
achieves a normal neutrino mass hierarchy $m_1\ll m_2\ll m_3$,
where the large atmospheric angle $\theta_{23}$ arises from the approximate equality of couplings $|e|\sim |f|$ of $\nu_R^{\mathrm{atm}}$ to 
$\nu_{\mu}$ and $\nu_{\tau}$, as in Eqs.~\ref{SRHN1}, \ref{SRHN2}.
The large solar angle $\theta_{12}$ arises from three roughly equal couplings $a \sim b \sim c$ of 
$\nu_R^{\mathrm{sol}}$ to $\nu_e$, $\nu_{\mu}$, $\nu_{\tau}$ in Eq.~\ref{schematic}. 
In this way the large mixing angles arise from ratios of couplings to the same right-handed neutrino and a normal neutrino mass hierarchy 
can coexist without relying on accidental cancellations.
Assuming $d\approx 0$ the reactor angle is given by $\theta_{13}\lesssim m_2/m_3$~\cite{King:2002nf} in agreement with the data. 

Motivated by the above results, maximal atmospheric mixing $\tan \theta_{23}= 1$ ($\theta_{23}=45^{\circ}$) suggests that $|e|= |f|$ and tri-maximal solar mixing
$\tan \theta_{12}= \frac{1}{\sqrt{2}}$ ($\theta_{12}=35.26^{\circ}$) suggests that $|b-c|= 2|a|$, while small reactor angle suggests a texture zero $d=0$, as mentioned above
~\footnote{Alternatively, golden ratio solar mixing may be considered with $\tan \theta_{12}= \frac{1}{\phi}$, where $\phi=(1+\sqrt{5})/2$ is the golden ratio ($\theta_{12}=31.7^{\circ}$)~\cite{Ding:2017hdv}. However, when corrections from the reactor angle are taken into account, this leads to a prediction for the solar angle outside the current 3$\sigma$ region~\cite{Costa:2023bxw}. For other alternative cases using CP symmetry, see~\cite{Ding:2018fyz,Ding:2018tuj}.}.
This motivates choosing $e=f$, $b=n a$ and $c=(n-2)a$ in Eq.~\ref{3by2}, called constrained dominance sequence (CSD) for some real number $n$~\cite{King:2005bj,Antusch:2011ic,King:2013iva,King:2015dvf,King:2016yvg,Ballett:2016yod,King:2018fqh,King:2013xba,King:2013hoa,Bjorkeroth:2014vha},\footnote{There is also a flipped case with the same first column $(0, e, e)^T$ and the second column $(a, (n-2)a, na)^T$ which 
yields related predictions $\theta_{23} \rightarrow \pi - \theta_{23}$ and $\delta \rightarrow \delta + \pi$.}
\begin{align}
	m^D=\left(\begin{array}{cc}
		0 & a \\
		e & n a \\
		e & (n-2) a
	\end{array}\right),
	\label{CSDn}
\end{align}
whereupon Eq.~\ref{2rhn} becomes,
\begin{equation}
m^{\nu}=
\frac{|e|^2}{M_{\rm atm}}\left( \begin{array}{ccc}
0 & 0 & 0 \\
0 & 1 & 1\\
0 & 1 & 1
\end{array}
\right) +
\frac{|a|^2}{M_{\rm sol}}e^{i\eta}
\left( \begin{array}{ccc}
1 & n & (n-2)  \\
n & n^2 & n(n-2) \\
(n-2)& n(n-2) & (n-2)^2
\end{array}
\right),
\label{2rhnCSD}
\end{equation}
where $\eta= 2\arg(a/e)$.
The choice $n\approx 3$ provides a particularly good fit to neutrino oscillation data and is called the Littlest Seesaw (LS) \cite{King:2015dvf}. 
For example models based on CSD($3$)~\cite{King:2013iva,King:2015dvf,King:2016yvg,Ballett:2016yod,King:2018fqh}, CSD($2.5$)~\cite{Chen:2019oey}
may arise from vacuum alignment, and CSD($1+\sqrt{6}$) $\approx$ CSD($3.45$)~\cite{Ding:2019gof,Ding:2021zbg,deMedeirosVarzielas:2022fbw,deMedeirosVarzielas:2023ujt,deAnda:2023udh}
from modular symmetry. For a given value of $n$, predictions for the PMNS matrix and the three neutrino masses can be analytically derived 
from the three real input parameters in Eq.~\ref{2rhnCSD}, 
namely $m_a= \frac{|e|^2}{M_{\mathrm{atm}}}$, $m_b=\frac{|a|^2}{M_{\mathrm{sol}}}$,
$\eta= 2\arg(a/e)$~\cite{King:2015dvf}. In practice, the three input parameters $m_a, m_b, \eta$ 
may be fixed using the three best measured observables $\theta_{13},\Delta_{31}^2,\Delta_{21}^2$, leading to genuine predictions
for the least well measured observables $\theta_{23}$, $\delta$ and $\theta_{12}$,
as shown in Table~\ref{LSpredictions}~\cite{Costa:2023bxw}. Note also that the neutrino mass matrix in Eq.~\ref{2rhnCSD}
implies that the first column of the PMNS matrix has the fixed magnitudes 
$(\sqrt{\frac{2}{3}}, \sqrt{\frac{1}{6}}, \sqrt{\frac{1}{6}})$
for any value of $n$, leading to atmospheric mixing sum rules which implies
$(\cos^2\theta_{12})(\cos^2\theta_{13})=2/3$ and predicts $\cos \delta$ in terms of the atmospheric and reactor angles~\cite{King:2015dvf}. The results in Tab.~\ref{LSpredictions}, are consistent with these atmospheric sum rule predictions, and fix the atmospheric angle, depending on $n$, for normal and flipped cases~\cite{Costa:2023bxw}.

\begin{table}[t]
	\centering
	\begin{tabular}{||c||c||c||c||c||}
		\hline \hline
		CSD($n$) & angle   &  $  n = 2.5$ &  $n=3$ & $  n = 3.45$ \\ \hline
				normal &$\theta_{23}\; [^{\circ}]$  & $51.5^{+1.9}_{-2.2} $ & $45.5{}^{+2.3}_{-2.4} $ & $41.4^{+2.6}_{-2.6}$\\ \hline
	    normal &	$\delta\; [^{\circ}]$ & $299.9^{+9.2}_{-9.9}$ & $272.2^{+9.6}_{-11.0}  $ & $253.8{}^{+11.7}_{-13.8}$ \\ \hline \hline
		flipped &$\theta_{23}\; [^{\circ}]$  & $38.5^{+1.9}_{-2.2} $ & $44.5{}^{+2.3}_{-2.4} $  & $48.6^{+2.6}_{-2.6}$ \\ \hline
        flipped &	$\delta\; [^{\circ}]$ & $119.9^{+9.2}_{-9.9} $ & $92.2^{+9.6}_{-11.0} $ & $74.8{}^{+11.7}_{-13.8}$ \\ \hline \hline
        both & $\theta_{12}\; [^{\circ}]$  & $34.31^{+0.16}_{-0.20} $  & $34.32{}^{+0.20}_{-0.24} $  & $34.36^{+0.18}_{-0.21} $ \\ \hline \hline
			\end{tabular}
	\caption{\footnotesize The CSD($n$) predictions for $\theta_{23}$, $\delta$ (for normal and flipped cases) and $\theta_{12}$
	(common to both cases), where the most accurately measured observables
	$\theta_{13},\Delta_{31}^2,\Delta_{21}^2$ are used to fix the three parameters 
	$ \eta,m_a, m_b$ leading to the uncertainties shown. Predictions are shown for three values of $n\approx 3$, arising from particular Littlest Seesaw models. Table adapted from~\cite{Costa:2023bxw}.}
	\label{LSpredictions}
\end{table}

The heavy-light mixing angles for SD are given from Eqs.\ref{theta}, \ref{eq:RH-masse}, \ref{3by2},
\begin{align}
	\theta_{lN}=\left(\begin{array}{cc}
		\frac{d}{M_{\mathrm{atm}}} & \frac{a}{M_{\mathrm{sol}}} \\
		\frac{e}{M_{\mathrm{atm}}}  &  \frac{b}{M_{\mathrm{sol}}} \\
		\frac{f}{M_{\mathrm{atm}}}  &  \frac{c}{M_{\mathrm{sol}}}
	\end{array}\right). 
	\label{theta3by2}
\end{align}
For CSD($3$) (normal case of Littlest Seesaw),
\begin{align}
		|\theta_{lN}|^2\approx \left(\begin{array}{cc}
		 0 &\frac{m_2}{3M_{\mathrm{sol}}} \\
		\frac{m_3}{2M_{\mathrm{atm}}} & \frac{3m_2}{M_{\mathrm{sol}}} \\
		\frac{m_3}{2M_{\mathrm{atm}}} & \frac{m_2}{3M_{\mathrm{sol}}}
	\end{array}\right)
	\approx \left(\begin{array}{cc}
		 0 &0.03\\
		0.25 &0.26\\
		0.25 &0.03
	\end{array}\right)\times10^{-10}\left( \frac{1 {\rm GeV}}{M}  \right),
	\label{thetasquared3by2}
\end{align}
where we have used Eqs.~\ref{masshierarchy},\ref{m2},\ref{m3},\ref{CSDn},\ref{theta3by2},
which may be compared to the single right-handed neutrino estimate in Eq.~\ref{theta1}, 
based on $|\theta|^2\approx m_{\nu}/M$ where a light neutrino mass of $m_{\nu}=0.1$ eV was taken.
The predictions in Eq.~\ref{thetasquared3by2} seem to be out of reach of the planned future experimental searches in Fig.~\ref{U2}.
In any case, for hierarchical right-handed neutrino masses, leptogenesis requires the lightest right-handed neutrino to have a mass around $10^{10}$ GeV~\cite{Hirsch:2001dg,King:2002qh,Antusch:2006cw,Bjorkeroth:2015tsa,King:2018fqh}.
In such high scale seesaw models, the heavier right-handed neutrino will be associated with the $B-L$ breaking scale, whose breaking leads to cosmic strings which can generate an observable gravitational wave signature~\cite{Dror:2019syi,Fu:2023nrn}. GUTs can also be probed in this way~\cite{Buchmuller:2019gfy,King:2020hyd,King:2021gmj,Fu:2022lrn,Dunsky:2021tih,Fu:2024rsm}.

\subsection{A Heavy Dirac Neutrino}

Let us now consider the case of two RHNs $\nu_{R1},\nu_{R2}$ which form an off-diagonal mass term~\cite{King:2002nf},
\begin{equation}
\overline{\nu_{R1}^c}M \nu_{R2} +\text{H.c.}
\label{SM3}
\end{equation}
The $2\times 2$ complex symmetric mass matrix $M_R$ has the form,
\begin{align}
	M_R=\left(\begin{array}{cc}
		0 & M \\
		M & 0
	\end{array}\right).	
	\label{DiracRHN}
\end{align}
The Majorana masses on the diagonal are simply assumed to be zero for now, but later on their absence will be enforced by a symmetry.

The mass term in Eq.~\ref{SM3}, where $\nu_{R1}^c$
is a left-handed antineutrino, may be compared to the Dirac mass of the electron in Eq.~\ref{Dirac},
and for this reason the two right-handed neutrinos may be regarded as a single four component Dirac neutrino,
\beq
N= 
\begin{pmatrix}
\nu_{1R}^c  \\
\nu_{2R}
\end{pmatrix}
\label{N2}
\eeq
with a heavy Dirac mass $M\overline{N}N$.
Unlike the Majorana spinor in Eq.~\ref{N}, the Dirac spinor above contains four independent degrees of freedom.

In the flavour basis the Lagrangian below the electroweak scale is,
\beq
{\cal L}^{\rm lepton}_{\rm mass}  = -\sum_{l=e,\mu , \tau} \overline{l}_L m_l l_{R} 
- \overline{\nu}_{lL}m^D_{l1}\nu_{R1}
- \overline{\nu}_{lL}m^D_{l2}\nu_{R2}
- \overline{\nu_{R1}^c}M \nu_{R2} +\text{H.c.}
\label{heavyDirac}
\eeq
where the columns of $m_D= (m^D_{1}, m^D_{2})$ are constrained by neutrino observables to be \cite{Gavela:2009cd},
\begin{eqnarray}
m^D_{l1} & = &  m^D_{1} \left( \sqrt{1-\rho }\ U^{\rm PMNS}_{l2} + \sqrt{1+\rho }\ U^{\rm PMNS}_{l3} \right)
\nonumber \\
m^D_{l2} & = &  m^D_{2} \left( \sqrt{1-\rho } \ U^{\rm PMNS}_{l2} - \sqrt{1+\rho }\ U^{\rm PMNS}_{l3} \right), 
\label{Belen}
\end{eqnarray}
assuming a normal neutrino mass squared ordering, where
\beq
\rho \approx  \frac{ m_3 - m_2}{ m_3 +m_2}\approx 0.71, \ \ \ \  \sqrt{1-\rho }\approx  0.54, \ \ \ \  \sqrt{1+\rho }\approx 1.3.
\eeq

The low energy neutrino mass matrix is given by the seesaw formula in Eq.~\ref{seesaw2},
leading to the low energy neutrino mass matrix,
\beq
m^{\nu}_{ll'}\approx \frac{m^D_{l1}m^D_{l'2}}{M} + \frac{m^D_{l2}m^D_{l'1}}{M}
\label{mnu1}
\eeq
and the light neutrino mass eigenvalues,
\beq
m_1 = 0, \ \ \ \ m_2 \approx \frac{m^D_{1}m^D_{2}}{M} (1- \rho), \ \ \ \ m_3 \approx \frac{m^D_{1}m^D_{2}}{M} (1+ \rho).
\label{lightnumasses}
\eeq
Thus $m^D_{1}$ and $m^D_{2}$ are not constrained individually, only their product,
\beq
\frac{m^D_{1}m^D_{2}}{M}\approx 0.03 \ {\rm eV}.
\label{product}
\eeq

The heavy-light mixing angles for the two right-handed neutrinos $\nu_{Ra}$ (where $a=1,2$) are given from Eq.~\ref{theta} as
\beq
\theta_{la} \approx  \frac{m^D_{la}}{M}.
\label{theta1}
\eeq
The Dirac neutrino in Eq.~\ref{N2}, has heavy-light mixing described by a sum of squared mixing angles,
\beq
|\theta_{l}|^2 \approx  |\theta_{l1}|^2+ |\theta_{l2}|^2.
\label{theta11}
\eeq

It is interesting to compare the qualitative behaviour of the neutrino masses in Eq.\ref{lightnumasses},
\beq
m^{\nu}\sim \frac{m^D_{1}m^D_{2}}{M}
\eeq
to that of the squared heavy-light mixing angle in Eq.~\ref{theta11}, using Eqs.~\ref{theta1} and~\ref{Belen},
\beq
\theta^2\sim  \frac{(m^D_{1})^2}{M^2}+  \frac{(m^D_{2})^2}{M^2}.
\label{theta2}
\eeq
Whereas the neutrino masses depend on the product $m^D_{1}m^D_{2}$, the heavy-light mixing angles depend on the 
sum of squares $(m^D_{1})^2+(m^D_{2})^2$. If the two parameters $m^D_{1},m^D_{2}$ are hierarchical, with one very much larger than the other one, then enhanced heavy-light mixing will result. For example suppose that $m^D_{1}\ll m^D_{2}$, with their product fixed by Eq.\ref{product} to give the correct light neutrino masses, then the heavy-light mixing angle will be dominated by the second term in Eq.\ref{theta2} and  become enhanced. This can lead to arbitrarily large heavy-light mixing angles, depending on the ratio $m^D_{2}/m^D_{1}$. This means that, for hierarchical $m^D_{1}\ll m^D_{2}$ and accessible values of $M$, the heavy Dirac neutrino could be discovered experimentally, as shown in Fig.\ref{U2}.
Since the RHN is a Dirac particle, this means that the HNL will also be approximately Dirac and 
hence it would not be detected via lepton number violating processes such as shown in 
Figs.~\ref{WW}, \ref{ndbc}.

The above scenario raises some theoretical questions which can be addressed in explicit models. As already remarked, the form of the off-diagonal right-handed neutrino mass matrix in Eq.~\ref{DiracRHN}, with zero diagonal Majorana masses needs to be justified. Also the assumed hierarchy of $m^D_{1}$ and $m^D_{2}$ looks very {\it ad hoc}. These questions could be addressed by defining the two RHNs to have opposite lepton number, which forces them to form a Dirac fermion, then the Yukawa coupling involving the RHN with the ``wrong'' lepton number would be forbidden. The idea is that the forbidden dimension-4 Yukawa coupling may take a small lepton number violating value, together with small dimension-3 Majorana mass term for the RHNs (both operators violating lepton number by two units). Precisely how to achieve this is a model dependent question.
In the following we discuss two models which achieve this in different ways: the type Ib seesaw model and the Majoron model.

\subsubsection{Type Ib Seesaw Model}
\begin{figure}[t]
	\centering
	\includegraphics[width=.5\textwidth]{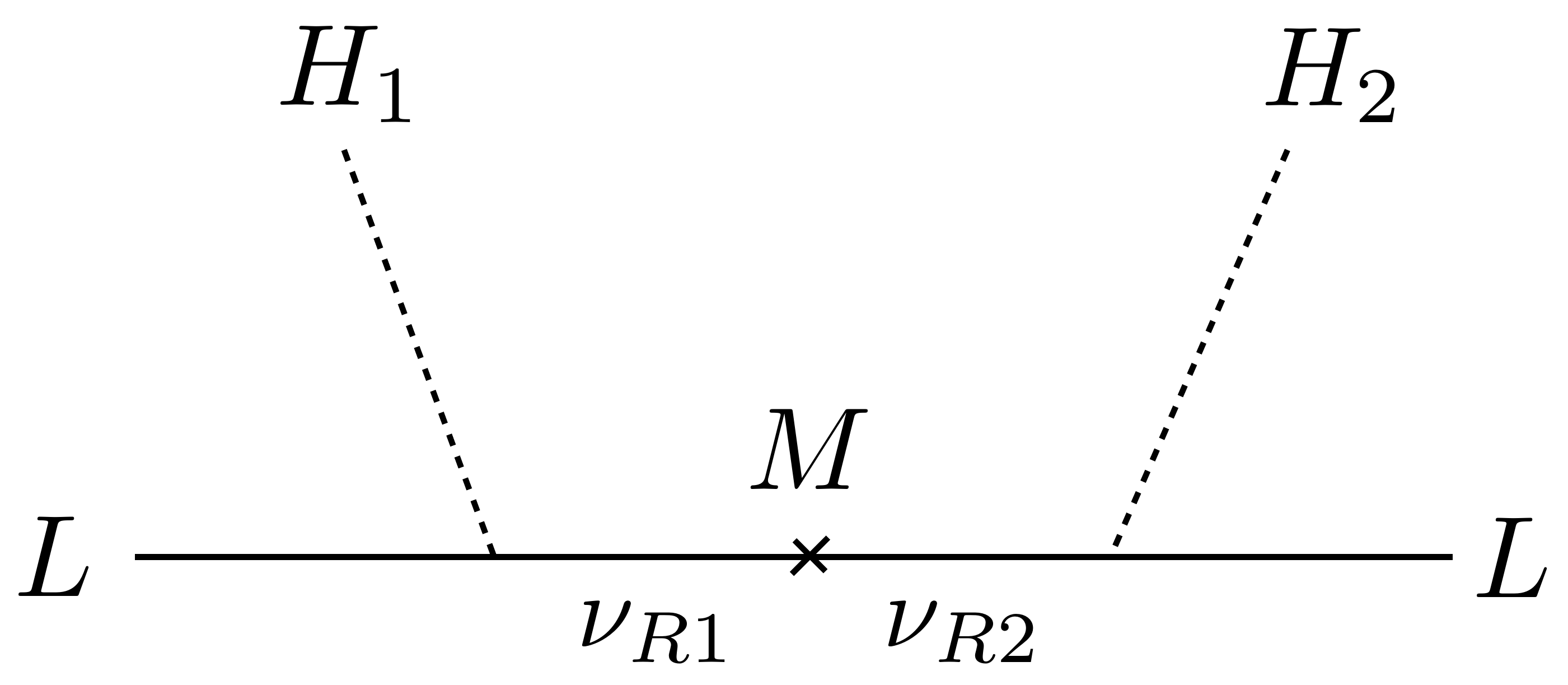}
	\caption{\footnotesize 
	The type Ib seesaw mechanism with two Higgs doublets $H_1$ and $H_2$ for a single heavy Dirac neutrino mass $M$.
	}
	\label{typeIb}
\end{figure}

In order to prevent diagonal Majorana masses in Eq.~\ref{DiracRHN} requires a symmetry under which the bilinear terms
$\nu_{R1}\nu_{R1}$ and $\nu_{R2}\nu_{R2}$ are forbidden while $\nu_{R1}\nu_{R2}$ is allowed. The simplest such symmetry is a discrete three-fold symmetry $Z_3$ symmetry (the rotational symmetry of an equilateral triangle) 
under which the right-handed neutrinos transform like $\nu_{R1}\rightarrow \omega^2 \nu_{R1}$ 
and $\nu_{R1}\rightarrow \omega \nu_{R1}$, where $\omega = e^{i2\pi/3}$ and hence $\omega^3=1$~\cite{Fu:2021fyk}.
However such a symmetry forbids the Yukawa couplings to right-handed neutrinos, $ \tilde{H}\overline{L}_i {y}^{\nu}_{ia} \nu_{Ra} $, and no single Higgs charge can allow couplings to both right-handed neutrinos. The solution is to introduce two Higgs doublets, $H_1$ with charge $\omega$ which couples to $\nu_{R1}$ and 
$H_2$ with charge $\omega^2$ which couples to $\nu_{R2}$, with the charged summarised in Table~\ref{typeIbcharges}.
The $Z_3$ symmetry assignments also ensure that the masses of the charged leptons and down-type quarks arise from the 
(CP conjugated) first Higgs doublet $\tilde{H}_1$, while the up-type quarks gain masses from $H_2$. This called a type II two Higgs doublet model~\cite{Branco:2011iw}. 
The lepton Lagrangian is given by~\cite{Fu:2021fyk},
\begin{equation}
{\cal L}^{\rm lepton}_{\rm mass}  = -\tilde{H}_1 \overline{L}_i {y}^e_{ij} e_{Rj} 
-{H_1} \overline{L}_i {y}^{\nu}_{i1} \nu_{R1}
 -{H_2} \overline{L}_i {y}^{\nu}_{i2} \nu_{R2}
- \overline{\nu_{R1}^c}M \nu_{R2} +\text{H.c.}
 \ ,
\label{LagrangiantypeIb}
\end{equation}
which yields the low energy Lagrangian in Eq.~\ref{heavyDirac}, now enforced by symmetry, with
\beq
m^D_{i1}= {y}^{\nu}_{i1}\frac{v_1}{\sqrt{2}}, \ \ \ \ m^D_{i2}= {y}^{\nu}_{i2}\frac{v_2}{\sqrt{2}},
\eeq
where $v_1, v_2$ are the VEVs of the two Higgs doublets $H_1, H_2$.
The neutrino masses then arise from a new form of seesaw mechanism which necessarily involves both Higgs doublets as in Fig.~\ref{typeIb}, and is known by the catchy name ``type Ib seesaw mechanism'' \cite{Hernandez-Garcia:2019uof,Chianese:2021toe,Fu:2021fyk}.
The low energy neutrino mass matrix is given by the seesaw formula in Eq.~\ref{seesaw2},
leading to the low energy neutrino mass matrix, as in Eq.~\ref{mnu1},
\beq
m^{\nu}_{ij}\approx \frac{m^D_{i1}m^D_{j2}}{M} + \frac{m^D_{i2}m^D_{j1}}{M}.
\label{mnu11}
\eeq

After integrating out the heavy Dirac neutrino, the type Ib seesaw mechanism leads to a new type of Weinberg
operator \cite{Hernandez-Garcia:2019uof},
\beq
\kappa_{ij} (L_i^T{H^*_1})(H_2^{\dagger}L_j),
\label{Wop2}
\eeq
which is similar to the original Weinberg operator in Eq.~\ref{Wop} but 
involving the two different Higgs doublets $H_1, H_2$.
Both Dark Matter  \cite{Chianese:2021toe} and Resonant Leptogenesis \cite{Fu:2021fyk} have been studied in the type Ib seesaw model. We emphasise that this model leads to a heavy Dirac neutrino as in Eq.~\ref{DiracRHN}, where the zero Majorana mass entries on the diagonal are not just assumed but are enforced by symmetry.

\begin{table}[t!]
\centering
\begin{tabular}{|c|c|c|c|c|c|c|c|c|c|}
\hline 
& ${Q}_i$ & ${u_R}_j$ & ${d_R}_j$
& $L_i$ & ${e_R}_j$ & $H_{1}$ & $H_{2}$  & $\nu_{\mathrm{R}1}$ & $\nu_{\mathrm{R}2}$  \\[1pt] \hline 
$SU(2)_L$ & {\bf 2} & {\bf 1} & {\bf 1} & {\bf 2} & {\bf 1} & {\bf 2} & {\bf 2} & {\bf 1} & {\bf 1} \\ \hline & & & & & & & & & \\ [-1em]
$U(1)_Y$ & $\frac{1}{6}$ & $\frac{2}{3}$ & $-\frac{1}{3}$ & $-\frac{1}{2}$ & $-1$ & $-\frac{1}{2}$ & $-\frac{1}{2}$ & 0 & 0  \\[2pt]  \hline & & & & & & & & & \\ [-1em]
$Z_3$ & $1$ &  $\omega$ &   $\omega$ & $1$ &  $\omega$ &   $\omega$ &   $\omega^2$ & $\omega^2$ &   $\omega$   \\ \hline  
\end{tabular}
\caption{\footnotesize \label{typeIbcharges}
A two Higgs doublet $H_1,H_2$ model under the electroweak $SU(2)_L\times U(1)_Y$ gauge symmetry and a discrete $Z_3$ symmetry (where $\omega =e^{i2\pi/3}$) \cite{Fu:2021fyk}. The fields $Q_{i}, L_{i}$ are left-handed SM doublets while ${u_R}_j,{d_R}_j,{e_R}_j$
are RH SM singlets where $i,j = 1,2,3$ label the three families of quarks and leptons
(see Eqs.~\ref{fam1}, \ref{fam2}, \ref{fam3}). 
The fields $\nu_{R1},\nu_{R2}$ are the two right-handed neutrinos.}
\end{table}

\subsubsection{Majoron Models}

We briefly mention another possibility for enforcing the off-diagonal right-handed neutrino masses in Eq.~\ref{DiracRHN},
which uses a Majoron field $\Phi$ whose VEV is responsible for their mass.
Since the Majorana masses of RHNs violate total lepton number $L$, their masses can arise from the spontaneous breaking of global $U(1)_{L}$ by a SM singlet complex scalar field $\Phi$, coupled to the RHNs, leading to a physical Goldstone boson called the Majoron~\cite{Chikashige:1980qk, Chikashige:1980ui, Gelmini:1980re}. 

Consider the charges shown in Table~\ref{Majoron} \cite{King:2024idj} where the Majoron field 
$\Phi$ is doubly charged under global lepton number $U(1)_L$ and is odd under a $Z_2$ symmetry, 
while the right-handed neutrinos have unit lepton number, with one even and one odd under $Z_2$.
All other SM fields (including the SM Higgs doublet $H$) are even under the $Z_2$, with leptons carrying the usual lepton number.

		\begin{table}[hbt]
		\begin{center}
			\vskip 0.5 cm
			\begin{tabular}{|c|c|c|c|}
				\hline
				Symmetries	&	$\Phi$ & $\nu_{R1}$ &  $\nu_{R2}$   \\
				\hline
				$U(1)_L$       &  $-2$     &  1 & 1  \\                 
				\hline
				$Z_2$    & $-$ &+ & $-$  \\                 
				\hline
			\end{tabular}
		\end{center}  
		\caption{\footnotesize Majoron field $\Phi$ and right-handed neutrino charges under global $U(1)_L$ and $Z_2$ symmetries.}
		\label{Majoron}
	\end{table}

The lepton Lagrangian, which respects both 
	$U(1)_L$ and $Z_2$ symmetry is given by,
	
	\begin{equation}
{\cal L}^{\rm lepton}_{\rm mass}  = -H \overline{L}_i {y}^e_{ij} e_{Rj} 
-\tilde{H} \overline{L}_i {y}^{\nu}_{i1} \nu_{R1} 
- \Phi \overline{\nu_{R1}^c}\nu_{R2} +\text{H.c.}
 \ .
\label{LagrangianMajoron}
\end{equation}
Note that lepton number $L$ forbids explicit right-handed neutrino masses, and these must be generated from the Majoron field couplings.
However the $Z_2$ assignments only allow one such Majoron coupling, namely the off-diagonal one, leading to the   
the off-diagonal right-handed neutrino masses in Eq.~\ref{DiracRHN}.
Note that $\nu_{R2}$ does not contribute to the renormalisable neutrino 
	Yukawa interaction since it is odd under $Z_2$. 
	However small soft $Z_2$ violating terms can provide the missing Yukawa coupling to $\nu_{R2}$, as well as an explanation for its smallness. Such terms will also lead to small Majorana mass terms, which splits the degeneracy of the two right-handed neutrinos, leading to resonant leptogenesis and Majoron dark matter  \cite{King:2024idj}. 
	
It is worth commenting that, without introducing Majoron fields, lepton number $L$, will enforce zero Majorana masses for all right-handed neutrinos, leading to only Dirac masses for the light neutrino fields. Such Dirac masses should be extremely small, namely around 0.1 eV or less, so, without a seesaw mechanism, the origin of such tiny neutrino masses would be very puzzling.
Of course the origin of the hierarchical quark and charged lepton masses is already puzzling, so this would simply add to the mystery. Since $L$ is a global symmetry, it is presumably softly broken at some level, so 
right-handed neutrino masses of some magnitude would seem to be generic. If combined with baryon number as 
$B-L$ then it becomes possible to gauge it, but then it should be spontaneously broken above the weak scale, so again right-handed neutrino masses would be expected. Nevertheless, Dirac physical neutrino mass models have been widely explored, and until the Majorana nature of light neutrinos has been established, notably by the observation of neutrinoless double beta decay,
they remain a possibility.

\section{Extra Singlet Neutrinos}
\label{extra}

\subsection{Double Seesaw vs Inverse Seesaw}

It is possible to introduce additional singlet fermions $S_R$ (which do not couple to the lepton doublets) in addition to the right-handed neutrinos $\nu_R$ 
(which have Dirac mass terms $m_D$). If the singlets have Majorana masses $\mu$, but the right-handed neutrinos
have zero Majorana masses $M_R=0$, then in the basis $(\nu_L^c,\nu_R,S_R)$, the mass matrix in block form is
\begin{equation}
\label{double}
\left( \begin{array}{ccc}
0& m_{D} & 0    \\
m_{D}^T & 0 & M \\
0 & M^T & \mu
\end{array}
\right)\,,
\end{equation}
which would be a $9\times 9$ matrix for three copies of each of $(\nu_L^c,\nu_R,S_R)$.

Assuming a hierarchy of the singlet mass matrix  $\mu$, the $(\nu_R,S_R)$ mass mixing matrix $M$, and the Dirac matrix $m_D$,
\beq
\mu \gg M\gg m_D,
\eeq
 then we have a two stage seesaw mechanism.
In the first stage the Majorana mass matrix $M_R$ is generated,
\begin{equation}
M_{R} = - M \mu^{-1}M^T.
\end{equation}
Then in the second stage the light physical left-handed
Majorana neutrino mass matrix is obtained in the usual way,
\begin{equation}
m_{\nu}= -m_{D}M_{R}^{-1}m_{D}^T \ .
\end{equation}
Combining these equations gives
\begin{equation}
\label{eq:inverse-seesaw}m_{\nu} = m_{D}{(M^{T})^{-1}} \mu M^{-1}m_{D}^T,
\end{equation}
which has a double suppression. This is called the double seesaw mechanism~\cite{Mohapatra:1986aw},
typically used in high scale models to explain why $M_{R}$ is below the GUT or string scale.
Perhaps surprisingly, the same formula, Eq.~\ref{eq:inverse-seesaw}, also applies to the case where $\mu$ is very small,
\beq
M\gg m_D\gg \mu\, ,
\eeq
which corresponds to the inverse seesaw mechanism
~\cite{Wyler:1982dd,Mohapatra:1986aw,Mohapatra:1986bd,Bernabeu:1987gr,GonzalezGarcia:1988rw,Akhmedov:1995vm,Akhmedov:1995ip,Malinsky:2005bi,Malinsky:2009df,Gavela:2009cd,Abada:2014vea}
\footnote{If one allows the 1-3 elements of Eq.~(\ref{double}) to be filled
in by a matrix $M'$~\cite{Akhmedov:1995vm} then one obtains another version of the low
energy seesaw mechanism called the linear seesaw mechanism~\cite{Akhmedov:1995ip,Akhmedov:1995vm,Malinsky:2005bi}.}.

\subsection{The Inverse Seesaw Mechanism}
To understand how the inverse seesaw mechanism works
let us first take the limit $\mu \rightarrow 0$, and consider one copy of $(\nu_L^c,\nu_R,S_R)$ (i.e. three chiral fermions). Eq.~\ref{double} with $\mu = 0$ then has the terms,
\beq
 \overline{\nu}_{R}(MS_{R}^c + m_D\nu_{L} ) = M  \overline{\nu}_{R}(S_{R}^c + \theta \nu_{L} )
 \equiv M  \overline{\nu}_{R}N,
\label{Diracheavy}
\eeq
where we have defined the small angle (c.f. Eq.~\ref{theta0}),
\beq
\theta = \frac{m_D}{M}.
\label{theta3}
\eeq
Eq.~\ref{Diracheavy} yields a single heavy Dirac neutrino of mass $M$ constructed from $\nu_R$ and $N$,
where the linear combination $N$ consists mainly of $S_{R}^c $ with a small admixture of $\nu_{L} $. The orthogonal linear combination 
$\nu$ is massless and consists mainly of $\nu_{L} $ with a small admixture of $S_{R}^c $,
\beq
N= S_{R}^c + \theta \nu_{L} , \ \ \ \ 
\nu =   \nu_{L} - \theta S_{R}^c 
\label{admixtures3}
\eeq
Inverting Eq.~\ref{admixtures3} gives,
\beq
S_{R}^c = N -  \theta \nu , \ \ \ \ 
\nu_{L}= \nu  +  \theta N,
\label{admixtures4}
\eeq
which may be compared to Eq.~\ref{admixtures2}. Similar to the usual seesaw, the heavy Dirac neutrino may be produced via its small admixture $\theta$ component of the active neutrino $\nu_L$.
The above analysis shows that the neutrino state $\nu$ (mainly $\nu_L$) is massless if $\mu=0$, and so lepton number $L$ is 
unbroken. Since the HNL in the inverse seesaw mechanism is a Dirac fermion, and $L$ is conserved to excellent approximation (see below), it would not be detected via the lepton number violating processes in 
Figs.~\ref{WW}, \ref{ndbc}.

Let us now include a small lepton number violating Majorana mass $\mu$ for the singlet $S_R$, 
\beq
\mu  \overline{S}_{R}S_{R}^c = \mu ( \overline{N^c} - \theta  \overline{\nu}^c  ) (N -  \theta \nu ) 
= \mu \theta^2 \overline{\nu}^c \nu + \cdots \rightarrow m_{\nu}=  \mu  \frac{m_D^2}{M^2} 
\label{inversemass}
\eeq 
which generates a small mass $m_{\nu}$ for $\nu$, using  Eq.~\ref{theta3}. 
It also generates a tiny mass splitting for the Dirac pair components $\nu_R$
and $N$, since the latter receives a negligible extra mass $\mu$. 
From  Eqs.~\ref{theta3}, \ref{inversemass}, 
\beq
|\theta|^2 \approx  \frac{|m_{\nu}|}{\mu},
\label{theta4}
\eeq
which is controlled by $\mu$ and is independent of the heavy neutrino mass $M$, unlike Eq.~\ref{theta0}.
This means that, for a fixed $m_{\nu}$, the heavy-light mixing angle is larger for small $\mu$.
For example, for $\mu = 1$ keV, and $m_{\nu}=0.1$ eV, the mixing angle would be $\theta= 10^{-2}$. This would be consistent with for example
$m_D=1$ GeV and $M=10^2$ GeV.

The above discussion may be readily generalised to the case of three active neutrinos $\nu_{Li}$, and three right-handed neutrinos $\nu_{Ra}$,
and three singlets $S_{Rb}$, in which case the inverse seesaw formula for the light neutrino mass matrix
takes the matrix form in Eq.~\ref{eq:inverse-seesaw}. In the limit that the mass matrix
$\mu \rightarrow 0$ three light active neutrinos remain massless and lepton numbers $L_e$, $L_{\mu}$, $L_{\tau}$
and total lepton number $L$ (the sum of the separate lepton numbers) 
are restored. The heavy-light mixing angles are calculated according to the matrix generalisation of Eq.~\ref{theta3},
\beq
\theta \approx  {m_D}M^{-1}.
\label{theta5}
\eeq
It is always possible to choose a basis where the heavy Dirac mass matrix $M$ is diagonal
by performing rotations on $\nu_R$ and $S_R$. However, having used up the transformations on $S_R$, the matrix $\mu$ will not be diagonal in general. This implies that the light neutrino mass matrix calculated using Eq.~\ref{eq:inverse-seesaw}, in the diagonal $M$ basis, will not only depend on the matrix $m_D$ but also on the arbitrary matrix $\mu$. This makes it very difficult in general to make any predictions for the neutrino masses and mixing angles. However it is possible to envisage a minimal case where predictivity is possible , as we now discuss.

\subsection{Minimal Inverse Seesaw Model}

The minimal example of the inverse seesaw mechanism involves two right-handed neutrinos and two singlets, so that 
the full mass matrix in Eq.\ref{double} has the form,
\begin{equation}
\left( 
\begin{array}{ccc}
0_{3\times 3} & (m_{D})_{3\times 2}  & 0_{3\times 2} \\ 
(m_{D}^{T})_{2\times 3}  & 0_{2\times 2} & M_{2\times 2}  \\ 
0_{2\times 3} & (M^{T})_{2\times 2}  & \mu_{2\times 2}  
\end{array}
\right) ,  \label{Mnu0}
\end{equation}
where $0_{n\times m}$ are $n\times m$ dimensional submatrices consisting of zero elements.
For example, one may achieve a natural mass hierarchy model based on sequential dominance (SD),
 with matrices as in Eqs.~\ref{eq:RH-masse}, \ref{3by2}, assuming that the matrix $\mu$ is diagonal
 in the same basis that $M$ is diagonal (a non-trivial assumption),
\begin{equation}
m_{D}= \left( 
\begin{array}{cc}
d & a \\
		e & b \\
		f & c
\end{array}
\right) ,\hspace{0.7cm}\hspace{0.7cm}
M= \left( 
\begin{array}{cc}
M_{\mathrm{atm}} & 0 \\
		0 & M_{\mathrm{sol}}
\end{array}%
\right) ,\hspace{0.7cm}\hspace{0.7cm}
\mu = \left( 
\begin{array}{cc}
\mu_{\mathrm{atm}} & 0 \\ 
0 & \mu_{\mathrm{sol}}
\end{array}%
\right).
\label{Mnublocks0}
\end{equation}%
Using the inverse seesaw formula in Eq.~\ref{eq:inverse-seesaw} with Eq.~\ref{Mnublocks0},
the light effective left-handed Majorana neutrino mass matrix
$m^{\nu}$ is now written as,
\begin{equation}
m^{\nu}=
\frac{ \mu_{\mathrm{atm}}}{M^2_{\rm atm}}\left( \begin{array}{ccc}
d^2 & de & df  \\
de & e^2 & ef\\
df & ef & f^2
\end{array}
\right)
+
\frac{ \mu_{\mathrm{sol}}}{M^2_{\rm sol}}\left( \begin{array}{ccc}
a^2 & ab & ac  \\
ab & b^2 & bc \\
ac & bc & c^2
\end{array}
\right),
\label{2rhnInverse}
\end{equation}
which has a similar form to the usual SD case in Eq.~\ref{2rhn},
where the SD condition in Eq.~\ref{SD1} replaced by,
\begin{align}
\frac{(e, f)^2\mu_{\mathrm{atm}}}{M^2_{\mathrm{atm}}}\gg 
	 \frac{(a, b, c)^2\mu_{\mathrm{sol}}}{M^2_{\mathrm{sol}}}.
	 \label{SD2}
\end{align}
The equations for the large solar and atmospheric mixing angles are unchanged from Eqs.~\ref{theta23}, \ref{theta12},
and the other results may be readily extended,
\begin{equation}
\tan\theta_{23}\approx\frac{|e|}{|f|}
\label{theta23ISS}
\end{equation}
\begin{equation}
\tan\theta_{12}\approx \frac{a}{\cos\theta_{23}b-\sin\theta_{23}c}
\label{theta12ISS}
\end{equation}
\begin{equation}
\theta_{13}\approx\frac{1}{m_{3}}
\left[
\frac{a\left(\sin\theta_{23}b+\cos\theta_{23}c\right)\mu_{\mathrm{sol}}}{M^2_{\mathrm{sol}}}
+\frac{d\sqrt{|e|^{2}+|f|^{2}}\mu_{\mathrm{atm}}}{M^2_{\mathrm{atm}}}
\right]
\end{equation}
\begin{equation}
m_{3}\approx\frac{(|e|^{2}+|f|^{2})\mu_{\mathrm{atm}}}{M^2_{\mathrm{atm}}}
\label{m3}
\end{equation}
\begin{equation}
m_{2}\approx \frac{a^{2}\mu_{\mathrm{sol}}}{M^2_{\mathrm{sol}}}
+\frac{(\cos\theta_{23}b-\sin\theta_{23}c)^{2}\mu_{\mathrm{sol}}}{M^2_{\mathrm{sol}}}, \ \ \ \ m_1\approx 0.
\label{m2}
\end{equation}
The heavy-light mixing angles are as in Eq.~\ref{theta3by2} but are now enhanced by having small $\mu$.
An inverse seesaw version of the Littlest Seesaw in Eq.~\ref{CSDn} with $n=3$
has also been considered leading to a predictive model~\cite{CarcamoHernandez:2019eme}.
The squared heavy-light mixing angles in this model are given by 
\begin{align}
		|\theta_{lN}|^2\approx \left(\begin{array}{cc}
		 0 &\frac{m_2}{3\mu_{\mathrm{sol}}} \\
		\frac{m_3}{2\mu_{\mathrm{atm}}} & \frac{3m_2}{\mu_{\mathrm{sol}}} \\
		\frac{m_3}{2\mu_{\mathrm{atm}}} & \frac{m_2}{3\mu_{\mathrm{sol}}}
	\end{array}\right)
		\label{thetasquared3by2inverse}
\end{align}
which differs from Eq.~\ref{thetasquared3by2} since here the angles are enhanced by small values of $\mu_{\mathrm{sol}}$,
 $\mu_{\mathrm{atm}}$, as in Eq.~\ref{theta4}. It is therefore entirely possible to have values of 
$M_{\mathrm{atm}}$ and $M_{\mathrm{sol}}$ in the GeV-TeV range, together with arbitrarily large
heavy-light mixing angles $|\theta_{lN}|^2$, allowing the inverse seesaw version of the SD model to be tested by experiment, remembering that the HNLs will be accurately Dirac in nature.

\section{Conclusion \label{sec:conc}}
The SM is a remarkably successful theory which accounts for all particle physics experimental results, except for neutrino mass and mixing, which requires it to be extended somehow. The non-renormalisable Weinberg operator can provide an origin for small neutrino masses and mixing, but it is not a complete theory,
and only provides an effective description of some unknown physics associated with some high energy scale 
$\Lambda$, which is responsible for the operator.

In this article we have discussed the simplest renormalisable extension of the SM capable of describing neutrino phenomenology,
consisting of the addition of right-handed neutrinos which are singlets under the SM gauge group and are sometimes referred to as sterile neutrinos, although they do have Yukawa couplings to active left-handed neutrinos.
Assuming a single right-handed neutrino with a Majorana mass much larger than its Dirac mass couplings to the left-handed active neutrinos, we introduced the seesaw mechanism, which provides an ultraviolet completion of the Weinberg operator and a natural suppression mechanism for the effective light left-handed Majorana neutrino masses that are observed in experiments. 
We also showed that while the predicted heavy neutral lepton could be experimentally observed if its mass was light enough,
its heavy-light mixing angle, which controls its coupling to $W$ bosons, is predicted to be too small, assuming one right-handed neutrino responsible for the atmospheric neutrino mass.

The addition of right-handed neutrinos to the SM thus not only provides a renormalisable origin of the Weinberg operator, but also comes with a prediction, namely the existence of heavy neutral leptons, corresponding to the extra degrees of freedom of the additional right-handed neutrinos to which they approximate. The masses of the heavy neutral leptons, which to good approximation may be regarded as the masses of the right-handed neutrinos, can range anywhere from the eV scale, to the keV scale, the GeV scale, the TeV scale and upwards to the GUT scale and beyond. They may show their presence virtually in loops which contribute for example to $\mu \rightarrow \gamma$, or if sufficiently light, may be produced as real states in current and planned particle physics experiments. However their visibility depends crucially on their couplings to heavy SM gauge bosons and to the Higgs, where such couplings depend on heavy-light mixing angles whose values depend on the unknown seesaw parameters.

The seesaw mechanism, like any renormalisable theory, involves more parameters than the Weinberg operators, namely the masses of the right-handed neutrinos, and the Dirac neutrino mass matrix elements in the flavour basis, which for example could be a $3\times 3$ complex matrix with 18 real parameters in the flavour basis, for the case of three right-handed neutrinos.
This sounds like a lot of parameters, but actually the same number of parameters also exists in the SM to describe each 
$3\times 3$ complex quark mass matrix. The difference is that many of these quark parameters are regarded as unphysical, since they cannot be measured by experiment, whereas the seesaw parameters are in principle physical measurable parameters, describing the heavy-light mixing angles responsible for PMNS unitarity violation and heavy neutral lepton couplings to SM particles.

The implementation of the seesaw mechanism is also quite model dependent. The starting point is the number of right-handed neutrinos that are added to the SM. The canonical choice is to add three right-handed neutrinos, one for each quark and lepton family, which is motivated by gauged $B-L$ models, and in particular $SO(10)$ GUTs where three right-handed neutrinos is a prediction. The seesaw mechanism with three right-handed neutrinos (below the $B-L$ breaking scale where neutrinos are SM gauge singlets) then involves an arbitrary Dirac mass matrix, as mentioned above, which may be parameterised in terms of the observable PMNS matrix (three angles plus three phases), the three light neutrino mass eigenvalues, the three right-handed neutrino masses, together with a complex orthogonal $3\times 3$ matrix which contains six real arbitrary parameters, making up the total of 18 real parameters of the Dirac mass matrix. Over some regions of parameter space the heavy neutral leptons may be light enough to produce in particle physics experiments with large enough heavy-light mixing angles to make them observable, and their presence may be indirectly be detected in lepton flavour violation and neutrinoless double beta decay.

Two right-handed neutrinos is the minimal choice consistent with current neutrino phenomenology, and this considerably reduces the number of seesaw parameters, with the complex orthogonal matrix now reducing to a $2\times 2$ matrix depending on one arbitrary complex angle. It also predicts that the lightest neutrino mass is zero, either in the normal or the inverted mass squared ordering case, which also removes one of the Majorana phases of the PMNS matrix. The two right-handed neutrino case could emerge as the limiting case of three right-handed neutrinos, where one of them makes a negligible contribution to the seesaw mechanism, and so is effectively decoupled. This is the case for sequential dominance, where one of the remaining two right-handed neutrinos gives the dominant contribution to the seesaw mechanism and is mainly responsible for the atmospheric neutrino mass splitting and mixing angle, while the other right-handed neutrino has a subdominant seesaw contribution and is mainly responsible for the solar mass splitting and mixing angle. Thus sequential dominance leads to a natural neutrino mass hierarchy together with large neutrino mixing angles, without any cancellations or tunings. This is a good starting point for predictions of the neutrino observables, as in the Littlest Seesaw Model, based on a version of constrained sequential dominance which can arise in theoretical models based on either vacuum alignment or modular symmetry, including $SO(10)$ models which typically lead to a strongly hierarchical right-handed neutrino mass matrix. However sequential dominance models share the single right-handed seesaw prediction of unobservably small heavy-light mixing angles 

Alternatively it is possible for the two right-handed neutrinos to have a large off-diagonal mass and to form a single heavy Dirac neutrino. In this case, the heavy-light mixing angle can become arbitrarily large, since one of the components of the heavy Dirac neutrino can have a large coupling to the left-handed active neutrinos, which is compensated by the small coupling of the other component, so that the light mass eigenstate, which depends on the product of these couplings, remains at its physical value.
From a theoretical point of view, it is necessary to explain why the diagonal heavy Majorana masses are forbidden, and why the couplings to the active neutrinos should be so asymmetric. We have discussed two such models which address these questions, the first one based on a two Higgs doublet model with a threefold discrete symmetry and a type Ib seesaw mechanism, the second one being a Majoron model with a twofold discrete symmetry and softly broken lepton number. 

It is also possible introduce extra singlet neutrinos which do not have any couplings to the active left-handed neutrinos but which do couple to right-handed neutrinos with zero Majorana masses.
If the extra singlet neutrinos have very large Majorana masses, they may generate effective Majorana masses for the right-handed neutrinos via a double seesaw mechanism. If the extra singlet neutrinos have very small masses, then they will control the masses of the lightest physical neutrino masses, via the same extended seesaw formula as for the double seesaw mechanism, although the physics is quite different as we discussed, and this case is called the Inverse Seesaw mechanism. The main phenomenological advantage of the Inverse Seesaw mechanism, is that the heavy-light mixing angles are controlled by the smallness of the extra singlet neutrino masses,
and so may be arbitrarily large leading to observable lepton flavour violation and possibly accessible 
neutral heavy Dirac leptons. With extra singlet neutrinos, the number of parameters grows, but the minimal Inverse Seesaw with two right-handed neutrinos plus two extra singlets, may lead to an inverse version of sequential dominance and the Littlest Inverse Seesaw model with the usual predictions for neutrino observables plus the added bonus of observable lepton flavour violation and heavy neutral leptons. 

Overall we have seen that the addition of right-handed neutrinos is a well motivated and minimal extension of the SM which is capable of accounting for all the neutrino observables, while providing the potentially testable prediction of heavy neutral leptons,
whose observability is highly model dependent. The freedom provided by the seesaw parameters leads to various theoretical models which have different features and predictive powers at low and high energy. The common feature of all these models is the seesaw explanation of the smallness of the neutrino masses as compared to the charged fermion masses. In addressing this problem, we are reminded that the charged fermions themselves are also subject to large mass hierarchies which are not addressed in the SM. This suggests that perhaps the seesaw mechanism could be extended in some way to address all the mass hierarchies of the SM. Such an extension would involve adding extra fermion states, which being charged would have to have large Dirac masses, and therefore be vector-like (i.e. non-chiral) under the SM gauge group. Like the seesaw mechanism, we should expect such extensions of the SM to involve more parameters, and yet provide a theoretically appealing explanation of the mass hierarchies of all the SM fermions, and, if we are lucky, provide a testable prediction of the experimental observation of the new heavy states that are introduced. Indeed such models have already been proposed, but that's another story.

\begin{ack}[Acknowledgments]%
The author thanks Martin Hirsch for discussions and 
acknowledges the STFC Consolidated Grant ST/X000583/1 and the European Union's Horizon 2020 Research and Innovation programme under the Marie Sklodowska-Curie grant agreement HIDDeN European ITN project (H2020-MSCA-ITN-2019//860881-HIDDeN).
\end{ack}



\end{document}